\documentclass[pre,showpacs,twocolumn]{revtex4}
\usepackage{graphicx}
\usepackage{natbib}
\usepackage{amsmath}

\begin{document}

\title{Cooperative Dynamics in a Network of Stochastic Elements with Delayed
Feedback}
\date{\today}
\author{D. Huber and L. S. Tsimring}
\affiliation{Institute for Nonlinear Science, University of California, 
San Diego, La Jolla, CA 92093-0402}

\begin{abstract}%
Networks of globally coupled, noise activated, bistable elements with
connection time delays are considered. The dynamics of these systems
is studied numerically using a Langevin description and
analytically using (1) a Gaussian approximation as well as (2) a
dichotomous model. The system demonstrates ordering phase transitions 
and multi-stability. That is, for a strong enough
feedback it exhibits nontrivial stationary states and oscillatory
states whose frequencies depend only on the mean of the time delay
distribution function.  Other observed dynamical phenomena include
coherence resonance and, in the case of non-uniform coupling
strengths, amplitude death and chaos. Furthermore, an increase of the
stability of the trivial equilibrium with increasing non-uniformity of
the time delays is observed.
\end{abstract}
 
\pacs{05.45.Xt, 05.40.Ca, 02.30.Ks, 02.50.Ey}
\maketitle

\section{Introduction}

Due to its relevance for a variety of scientific disciplines such as
physics, chemistry, biology, economics and social sciences, the
study of collective phenomena in extended stochastic systems with long
range interaction has been of great interest in recent years and
various techniques based on Langevin, Fokker-Planck and master
equations have been conceived to explore their dynamics.

An effective and simple model for the study of noise induced
collective phenomena is the globally coupled network of stochastically
driven bistable elements. Indeed, the cooperative dynamics of these
systems has been the subject of many studies and its relevance for
critical phenomena
\citep{Dawson83}, spin systems \citep{Jung92}, neural networks 
\citep{Koulakov02,Camperi98,Sompolinsky86}, 
genetic regulatory networks \citep{Gardner00} and decision making
processes in social systems \citep{Zanette97} has been pointed out.
 
For the sake of simplicity it has traditionally been assumed that the
interactions in these networks are instantaneous. However, in recent
years it has been realized that time delays due to finite
transmission and processing speeds are 1) significant compared to the
dynamical time scales of the system and 2) often change fundamentally
its dynamical properties \citep{Reddy98,Nakamura94,Bresseloff98,Choi85}.

Thus, in this paper the generic model of globally coupled bistable
elements is extended by time delayed couplings and its collective
dynamics is studied numerically and analytically.

The properties of globally coupled dynamical units, relevant
for system such as arrays of lasers \citep{Wiesenfeld90} and Josephson
junctions \citep{Wiesenfeld89}, have been explored in many studies
\citep[see also][]{Yeung99,Broeck94,Jung92,Kuramoto91,Shiino87,Desai78}.
\citet{Desai78}, for instance, studied the synchronization of noise activated
bistable oscillators with instantaneous coupling and derived from the
Fokker-Planck equation for the joint probability distribution of the
oscillators an exact mean field model (DZ-model) in the thermodynamic
limit $N\to\infty$, where $N$ is the number of network
elements. Beyond a critical coupling strength this system displays a
second order phase transition to an ordered nontrivial stationary
state. The effect of uniform interaction delays in a globally coupled
network of phase oscillators has been explored by
\citet{Yeung99}, and \citet{Tsimring01} studied the dynamics of a single
noise activated bistable element with delayed feedback. Combining the
properties of these two systems, \citet{Huber03} studied the properties
of a globally coupled network of noisy bistable elements with uniform
delays and derived a dichotomous mean field model based on the delay-differential 
master equation. Although for numerous systems the
assumption of uniform time delays is justified
\citep[e.g.][]{Salami03,Paulsson01}, most systems have time delays distributed
over an interval rather than concentrated at a point
\citep[see][]{Atay03}. Thus, after a discussion of the dynamical
properties of the network with uniform delays, the generalized case of
distributed time delays is studied.

This paper which is an extended version of \cite{Huber03}
is organized as follows: In the next Section the
bistable-element-network is discussed for the case of uniform time
delays. Two mean field models, namely the DZ-model (which for Gaussian
processes reduces to the Gaussian approximation) and the dichotomous
model, are compared with the Langevin dynamics and their scopes of
application are determined. The phenomenon of coherence resonance is
discussed and a complete bifurcation analysis of the trivial
equilibrium is carried out using a center manifold reduction. In
Sect.~\ref{twodelays} the system dynamics is discussed for a discrete
bimodal delay distribution. Then, in Sect.~\ref{multidelays} the model
is further generalized, so that the mean field dynamics 
of a system with an arbitrary time delay distribution can
be described. Finally, in Sect.~\ref{twocouplings} we introduce
nonuniform coupling strengths which lead to new dynamical properties,
such as amplitude death and chaos.

\section{Uniform delays}
\label{uniform}

\subsection{Langevin model}

The prototypical system considered here is modeled by a set of $N$
Langevin equations, each describing the overdamped stochastically
driven motion of a particle in a bistable potential $V=-x^2/2+x^4/4$,
whose symmetry is distorted by the time delayed coupling to all
network elements,
\begin{equation}
\label{languniform}
\dot{x}_i(t)= x_i(t)-x_i(t)^3+\frac{\varepsilon}{N} 
              \sum_{j=1}^N x_j(t-\tau)+\sqrt{2D}\xi(t),
\end{equation}
where $\tau$ is the time delay, $\varepsilon$ is the coupling strength
of the feedback and $D$ denotes the variance of the Gaussian
fluctuations $\xi(t)$, which are mutually independent and uncorrelated
$\langle \xi_i(t)\xi_j(t') \rangle = \delta_{ij}\delta(t-t')$.

The global coupling leads to an asymmetry of the local potential; that
is, a positive feedback increases the probability for an element to be in
 the potential well in which the majority of elements were at
time $t-\tau$. The inverse holds for a negative feedback.

System (\ref{languniform}) is explored numerically. In this paper,
the numerical simulations are carried out using an Euler method. If not
otherwise indicated the time-step and the number of elements are
$\Delta t=0.01-0.05$ and $N=2500$.

Our interest is mainly focused on the cooperative interactions of the
individual network elements, i.e., on the dynamics of the mean field
$X=N^{-1}\sum_{j=1}^N x_j$. For $\varepsilon=0$, the elements are
decoupled from each other. They jump from one potential well to the
other randomly and independently of each other. Therefore, in this
case the mean field $X=0$. For small $|\varepsilon|$, the mean field
remains zero. At a certain $\varepsilon=\varepsilon_{\rm st}>0$ which
depends on the noise intensity $D$, but is independent of the time
delay, the system undergoes a second order (continuous) phase
transition and adopts a non-zero stationary mean field.

For a negative feedback, a transition to a periodically oscillating
mean field solution is observed at a certain
$\varepsilon=\varepsilon_{\rm osc-}<0$. Here and for the rest of this
paper a $(-/+)$ index means that the corresponding value is associated
with a negative/positive feedback.

Above a certain noise level $D_{H}$ the transition at
$\varepsilon_{\rm osc-}$ is second order as well. However, for
$D<D_{H}$ the system exhibits a first order (discontinuous) transition
associated with hysteretic behavior. The noise intensity $D_{H}$
depends on the time delay and is $D_{H}=0.07$ for $\tau=100$.

For large time delays $\tau\gg\tau_K$ ($\tau_K$ is the inverse Kramers
escape rate from one well into the other \citep{Haenggi90,Kramers40}),
depending on the initial state the system adopts one of many
accessible oscillatory states featuring different periods.  Even for a
positive feedback, besides the stationary solution several oscillatory
states with periods $T\lesssim\tau$ are observed for
$\varepsilon>\varepsilon_{\rm osc+}\gtrsim\varepsilon_{\rm st}$.  If
the feedback is negative, the system only has oscillatory nontrivial
solutions. The observed periods are $T\lesssim2\tau$ for
$\varepsilon<\varepsilon_{\rm osc-}$.

\begin{figure*}
\includegraphics[angle=-90,width=16cm]{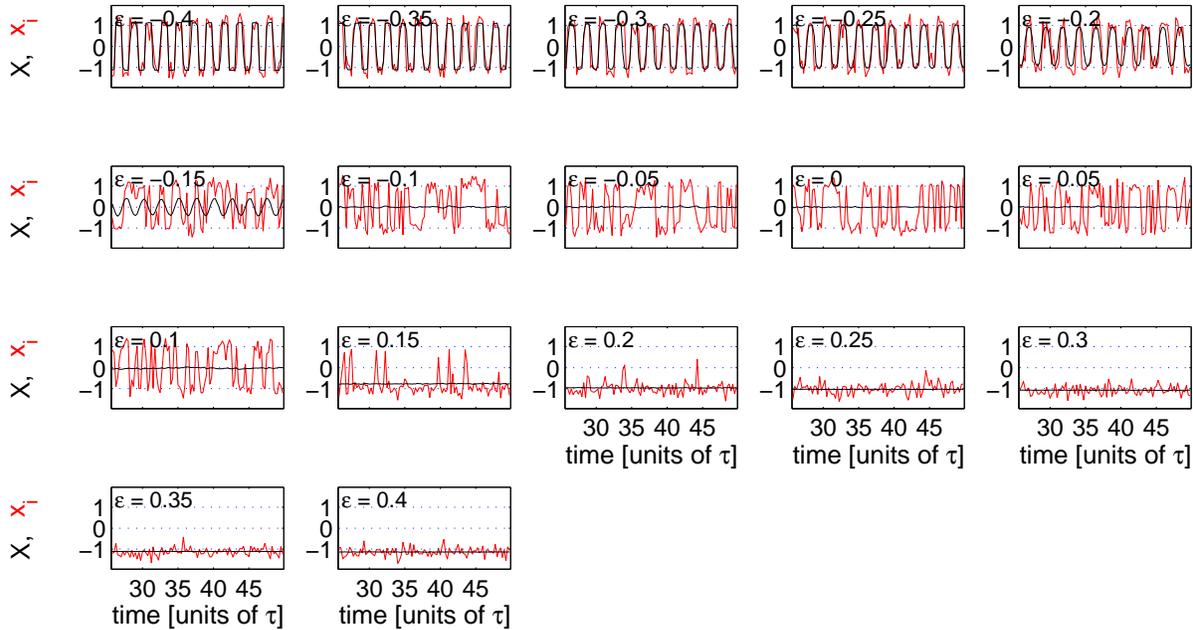}
\caption{\label{hyst} Dynamics of a single network oscillator
$x_i$ and the network mean field $X$ for different coupling strength
$\varepsilon$. The noise strength and the time delay is $D=0.1$ and
$\tau=100$, respectively. For $\varepsilon<\varepsilon_{\rm osc}=-0.13$
the mean field adopts a state of periodic oscillations. In the range
$\varepsilon_{\rm osc} <\varepsilon<\varepsilon_{\rm st}=0.11$ the
trivial equilibrium is stable.  Finally, for
$\varepsilon>\varepsilon_{\rm st}$ the system adopts a non-zero
stationary state.}
\end{figure*}
\begin{figure*}
\includegraphics[angle=-90,width=16cm]{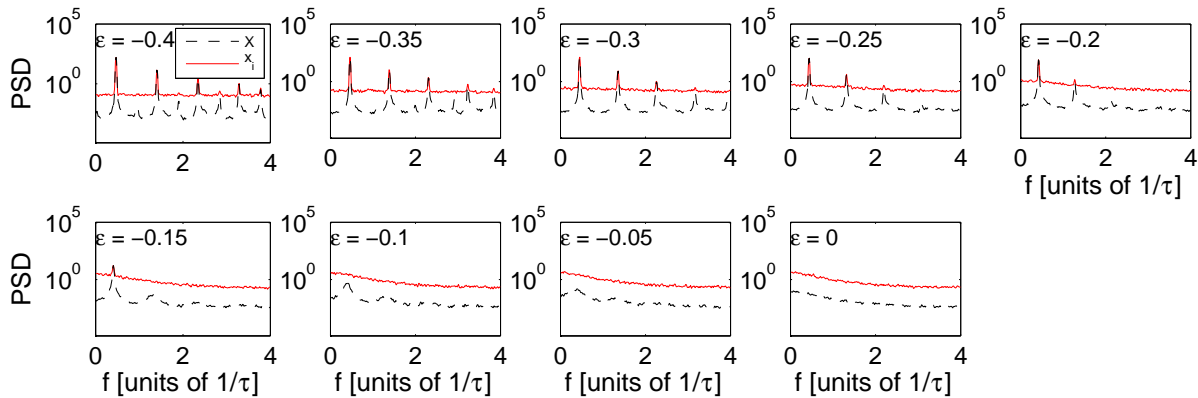}
\caption{\label{fhyst} For $\varepsilon\le0$, same as in Fig. 
\ref{hyst}, but in the frequency domain.}
\end{figure*}

The basic dynamical states accessible by the system are illustrated in
Figs.~\ref{hyst} and \ref{fhyst}, where the evolution of a single
network element and the mean field are shown for different coupling
strength.

\subsection{Gaussian approximation}

A mean field description for the dynamics of a globally coupled set of
thermally activated bistable elements with instantaneous interactions
was proposed by \citet{Desai78}. For the sake of simplicity, we refer
to this mean field description as the DZ-model. This model consists of
a hierarchy of equations for the cumulants of the distribution
function derived from the Fokker-Planck equation for the joint
probability density function of all elements. Expressed in terms of
moments $M_n\;\{n=1\ldots\infty\}$ the hierarchy assumes the simple
form,
\begin{equation}
\dot M_n=X(t-\tau)[4DM_{n-2}+\varepsilon M_{n-1}]+M_n-M_{n+2},
\end{equation}
where $M_{-1}=0$ and $M_0=1$. 

For large noise intensities, when the statistics of the individual
elements are approximately Gaussian the hierarchy can be truncated
(Gaussian approximation \citep[see also][]{Pikovsky02}).  Applying
this approach to our delayed feedback system, the evolution of the
mean field $X$ is, in the Gaussian approximation, described by the
following set of equations,\\
\parbox{7.5cm}{
\begin{eqnarray*}
\dot{X}(t)&=& X(t)-X^3(t)-3X(t)V(t)+\varepsilon X(t-\tau),
\label{XM}\\
\frac{1}{2}\dot{V}(t)&=& V(t)-3X^2(t)V(t)-3V^2(t)+D,
\end{eqnarray*}
}\hfill
\parbox{8mm}{\begin{eqnarray}\end{eqnarray}}
where $V=M_2-M^2_1=N^{-1}\sum_{i=1}^N(x_i-X)^2$ is the variance. 

\begin{figure}
\includegraphics[width=8cm]{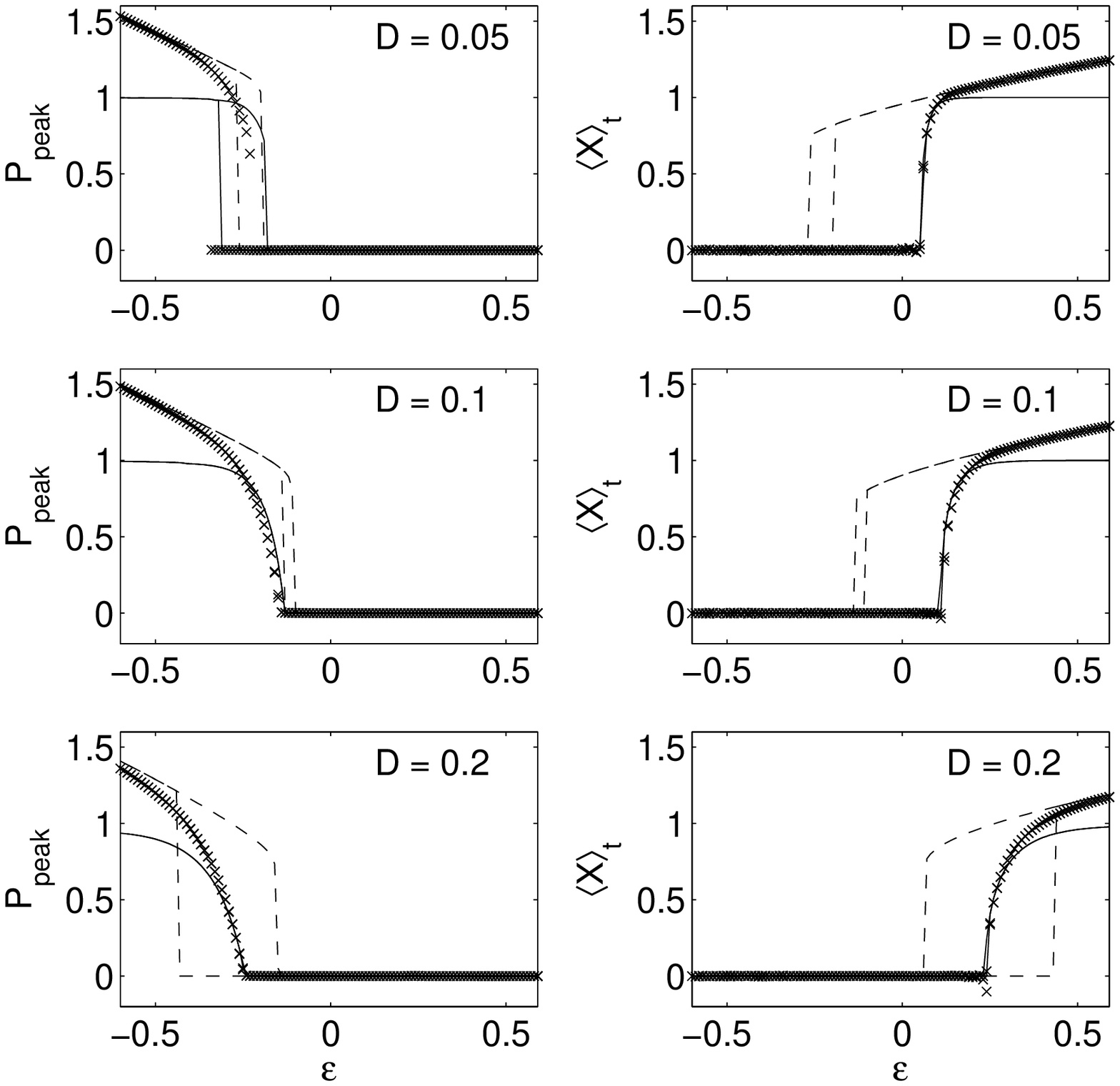}
\caption{\label{mixall} The peak power $P_{\rm peak}$ and the temporal
mean $\langle X \rangle_t$ as a function of the coupling strength for
the Langevin model (crosses), the Gaussian approximation (dashed
line), where the double line indicates hysteretic behavior, and the
dichotomous theory (solid line). The noise strengths is indicated in
the upper right corner of each panel. The time delay is
$\tau=100$. For $X=0$ and $D=0.05\,,0.1\,,0.2$ the Kramers times are
$\tau_K=659.4\,,54.1\,,15.5$.}
\end{figure}

To compare the theoretical predictions of the Gaussian approximation
(\ref{XM}) with the Langevin model (\ref{languniform}) we determine
the maximum of the main peak in the power spectrum $P_{\rm peak}$ (see
Fig. \ref{fhyst}).  The evolution of the peak power as a function of
the coupling strength can be used to study the Hopf bifurcation which
describes the transition to the oscillatory mean field regime. The
pitchfork bifurcation describing the transition to the stationary mean
field state is characterized by the dependence of the temporal mean of
the mean field $\langle X\rangle_t$ on the coupling strength.
Fig.~\ref{mixall} shows the peak power $P_{\rm peak}$ and the temporal
mean $\langle X\rangle_t$ as a function of the coupling strength
$\varepsilon$ for three different noise temperatures $D$.

\begin{figure}
\includegraphics[angle=-90,width=8cm]{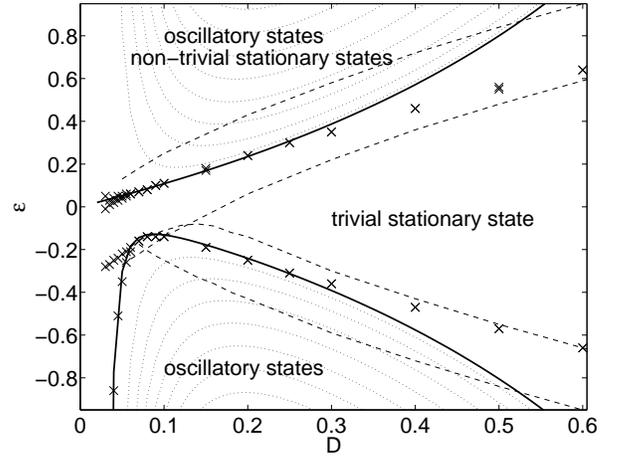}
\caption{\label{phase} Phase diagram for $\tau=100$ of the 
Langevin model (crosses), the Gaussian approximation (dashed lines)
and the dichotomous theory (solid lines and dotted lines).  The solid
line and the dotted line respectively depict the primary solution and
the higher order solutions of Eq.~(\ref{omegaeps1}) and
(\ref{omegaeps2}). Phases separated by double lines indicate
hysteretic behavior. For $X=0$ and $D<0.3$ the Kramers time is
$\tau_K>10$.}
\end{figure}

The phase diagrams of these models are shown in Fig.~\ref{phase} in the
$(D,\varepsilon)$-parameter plane. Fig.~\ref{mixall} shows that away
from the transition points the Gaussian approximation correctly
describes the Langevin dynamics. However, near the bifurcation points
the system dynamics is strongly non-Gaussian even for strong noise.
Indeed, while the Gaussian approximation predicts that both
bifurcations are first order transitions (associated with hysteretic
behavior) over the entire noise range considered in this study (see
Fig.~\ref{phase}), the Langevin model produces first order
transitions only for $D<0.7$.
 
\begin{figure}
\includegraphics[width=5cm]{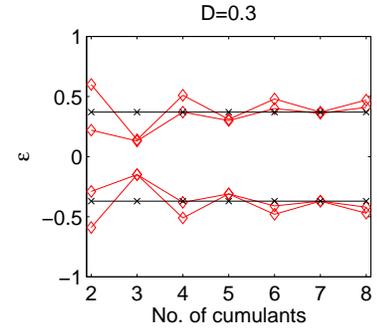}
\caption{\label{varord} The critical coupling of the Hopf bifurcation
$(\varepsilon_{\rm osc}^1<0)$ and the pitchfork bifurcation
$(\varepsilon_{\rm st}>0)$, respectively. Compared are the critical
couplings resulting from the Langevin model (crosses) and the
predictions of the DZ-theory (diamonds) including different numbers of
cumulants. For an even number of cumulants the DZ-theory predicts
hystertic behavior which is not seen in the Langevin dynamics.}
\end{figure}

The inclusion of higher order cumulant equations leads only to a slow
convergence toward the true solution of the Langevin model. This is
illustrated in Fig.~\ref{varord}. Thus, near the transition points the
DZ-model does not significantly simplify the Langevin description and
the critical parameters for the transition cannot be determined
analytically.

\subsection{Dichotomous Model}

\label{dimod1}
In order to describe the dynamics of the system near the bifurcation
points we apply a {\it dichotomous} (i.e. two-state) approximation,
which is complementary to the Gaussian approximation and which is
valid in the limit of small noise, when the characteristic Kramers
transition time is $\tau_K\gg 1$, and small coupling strengths. The
dichotomous theory neglects intra-well fluctuation of $x_i$. Thus, in
the limit of small coupling, each bistable element can only take the
values $s_{1,2}=\pm 1$. The collective dynamics of the entire network
can then be described by the master equations for the occupation
probabilities of these states $n_{1,2}$. This approach has been
successfully used in studies of stochastic and coherence resonance
\citep[e.g.][]{McNamara89,Gammaitoni98,Jung92,Tsimring01}. For example,
using this approach \citet{Jung92}, found nontrivial stationary mean
field solutions in a globally coupled delay-free network of bistable
elements.

The dynamics of a single element is determined by the hopping
rates $p_{12}$ and $p_{21}$, i.e., by the probabilities to hop over
the potential barrier from $s_1$ to $s_2$ and from $s_2$ to $s_1$,
respectively.  In a globally coupled
system, $n_{1,2}$ and $p_{12,21}$ are identical for all elements
and the master equations for the occupation probabilities read,
\begin{eqnarray}
\label{ndot1}
\dot{n}_1  &=&  -p_{12}n_1+p_{21}n_2\\
\dot{n}_2  &=&  \;\;\,p_{12}n_1-p_{21}n_2.
\end{eqnarray}
The hopping probabilities $p_{12,21}$ are given by Kramers' transition
rate \citep{Kramers40} for the instantaneous potential well,
\begin{equation}
p^{12,21}_{\rm K}=\frac{\sqrt{U''(x_m)U''(x_0)}}{2\pi}
\exp\left(\frac{-\Delta U}{D}\right)\;, 
\end{equation}
where $x_m$ and $x_0$ are the positions of the potential minima and
the top of the potential barrier, respectively.  For our system, in
the limit of small noise $D$ and coupling strength $\varepsilon$, they
read \citep[cf.][]{Tsimring01},
\begin{equation}
\label{hopuniform}
p_{12,21}=\frac{\sqrt{2\mp 3\alpha_1}}{2\pi}
\exp\left(-\frac{1\mp4\alpha_1}{4 D}\right),
\end{equation} 
where $\alpha_1=\varepsilon X(t-\tau)$.

As discussed above, the Langevin system either adopts a stationary or
an oscillatory mean field state in the limit $t\to\infty$.  Let us
first consider the stationary case $\dot{n}_{1,2}=0$. Making use of
the probability conservation $n_1+n_2=1$, the occupational
probabilities $n_{1,2}$ are given by
\begin{equation}
n_{1,2}=\frac{p_{21,12}}{p_{12}+p_{21}}\;.
\label{n12}
\end{equation}
Then, in the dichotomous approximation with $s_{1,2}=\pm 1$,
the mean field  $X=s_1 n_1+s_2 n_2$ reads
\begin{equation}
X=n_2-n_1=\frac{p_{12}-p_{21}}{p_{12}+p_{21}}\;.
\label{X0}
\end{equation}

Substituting the hopping probabilities ($\ref{hopuniform}$) into this 
equation yields the transcendental equation for the mean field magnitude
\begin{equation}
X=\frac{\sqrt{2-3\varepsilon X}\exp\left(\varepsilon X/D\right)
-\sqrt{2+3\varepsilon X}\exp\left(-\varepsilon X/D\right)}
{\sqrt{2-3\varepsilon X}\exp\left(\varepsilon X/D\right)
+\sqrt{2+3\varepsilon X}\exp\left(-\varepsilon X/D\right)}
\label{XX}
\end{equation}
This equation always has a trivial solution $X=0$, but for
$\varepsilon>\varepsilon_{\rm st}$ it also has a pair of nontrivial
solutions $X=\pm A$. It is easy to find $A(\varepsilon)$ for a fixed
$D$ numerically using (\ref{XX}).  The critical value
$\varepsilon_{\rm st}$ as a function of $D$ for the pitchfork
bifurcation, indicating the transition to a nontrivial stationary
state, can be found analytically by expanding the r.h.s. of
Eq.~(\ref{XX}) at small $X$.  This yields the following expression
\begin{equation}
\varepsilon_{\rm st}=\frac{4D}{4-3D}\;.
\label{ec}
\end{equation}

Let us now turn to the general case when $X$ is allowed to be a function
of time. Again, making use of the probability
conservation and the expression for the dichotomous mean field
$X=n_2-n_1$ we find the equation,
\begin{equation}
\dot X(t) = p_{12}-p_{21}-(p_{21}+p_{12})X(t),
\label{XXuniform}
\end{equation}
where the hopping probabilities $p_{12,21}$ have the same functional
form as in Eq.~(\ref{hopuniform}), but now depend on the delayed mean
field $X(t-\tau)$ rather than $X(t)$.

To investigate the stability properties of the system,
Eq.~(\ref{XXuniform}) is linearized about zero,
 \begin{equation}
\dot X(t)=\frac{\sqrt{2}}{\pi}\exp(-1/4D)
\label{lindich}
\left(\varepsilon\frac{4-3D}{4D}X(t-\tau)-X(t)\right)\;.
\end{equation}
The characteristic equation for the complex eigenvalue $\lambda$ is
found by making the ansatz $X\propto \exp(\lambda t)$. It reads,
\begin{equation}
\lambda=\frac{\sqrt{2}}{\pi}{\rm e}^{-1/4D}
\left(\frac{\varepsilon(4-3D)}{4D}{\rm e}^{-\lambda\tau}-1\right).
\label{lambdauniform}
\end{equation} 

The trivial equilibrium loses its stability and undergoes
a Hopf bifurcation indicating the transition to an oscillatory mean
field state when the real part of the complex eigenvalue becomes
positive. Therefore, the properties of the corresponding instabilities
(i.e. frequencies and coupling strengths at the bifurcation points) can be
found by substituting $\lambda=\mu+{\rm i}\omega$ into
Eq.~(\ref{lambdauniform}), separating real and imaginary parts and setting
$\mu=0$. This yields the following set of equations,
\begin{eqnarray}
\label{omegaeps1}
\omega\tau &=&-\frac{\sqrt{2}}{\pi}\exp(-1/4D)\tau\tan\omega\tau\\
\label{omegaeps2}
\varepsilon_{\rm osc}&=&\frac{\varepsilon_{\rm st}}{\cos\omega\tau}.
\end{eqnarray}
This set of equations has a multiplicity of solutions, indicating that
multi-stability occurs in the globally coupled system beyond a certain
coupling strength. For finite time delays and positive coupling,
besides the stationary solution, several oscillatory states with
periods $T_k$ close to but slightly larger than $\tau/k$ exist
for $\varepsilon>\varepsilon^{k}_{\rm osc+}\;\;\{k=1,2,\ldots\}$,
where the transition points are ordered as follows,
$0<\varepsilon_{\rm st}<\varepsilon^{1}_{\rm
osc+}<\varepsilon^{2}_{\rm osc+}\ldots$ If the feedback is negative,
the system has oscillatory solutions with periods $T_l$ close to but
slightly larger than $2\tau/(2l+1)$ for
$\varepsilon<\varepsilon^{l}_{\rm osc-}\;\;\{l=0,1,\ldots\}$, where
$0>\varepsilon^{0}_{\rm osc-}>\varepsilon^{1}_{\rm osc-}\ldots$ In the
limit of large time delays $\tau\to\infty$, the transition points
$\varepsilon^{k}_{\rm osc+}\to\varepsilon_{\rm st}$ and
$\varepsilon^{l}_{\rm osc-}\to\varepsilon^{0}_{\rm
osc-}=-\frac{4D}{4-3D}$ with the corresponding periods being
$T_{k}\to\tau/k$ and $T_{l}\to2\tau/(2l+1)$, respectively.

In order to compare the predictions of the dichotomous model with
the Langevin dynamics, the
peak power $P_{\rm peak}$ and the temporal mean $\langle X \rangle_t$,
resulting from the dichotomous theory, are also plotted in
Fig.~\ref{mixall}. The phase diagram for the dichotomous theory is
shown in Fig.~\ref{phase}.
  
Fig.~\ref{mixall} and \ref{phase} show that the dichotomous theory
agrees with the Langevin dynamics quite well for small noise in the
range $D\approx0.07-0.3$ in the neighborhood of the bifurcation
points. The theory also correctly describes the bifurcation type.
Indeed, the dichotomous theory predicts accurately the noise strength
$D_{\rm H}$ ($=0.07$ for $\tau=100$) at which the Hopf bifurcation
changes from supercritical (second order) to subcritical (first
order). However, for very small $D$ the Kramers time becomes very
large, and the accuracy of numerics becomes insufficient for a
comparison with the theory.

\subsection{Complete bifurcation analysis}

A complete bifurcation analysis of the trivial solution $X=0$ of
Eq.~(\ref{lindich}) in the $(D,\varepsilon,\tau)$-parameter space can
be accomplished by carrying out a center manifold reduction
\citep[see e.g.][]{Chow82,Faria95}; that is, the normal form coefficients
of the bifurcations in our dichotomous mean field model can be
expressed in terms of the system parameters.

For a general class of
delay differential equations of the form,
\begin{eqnarray}
\dot x(t)&=&x(t)+\gamma_1 x(t-\tau)+\gamma_2 x(t)^3 
+ \gamma_3 x(t)^2 x(t-\tau)\nonumber\\
         & & +\gamma_4 x(t)x(t-\tau)^2+\gamma_5x(t-\tau)^3,
\label{manif}
\end{eqnarray} 
such a reduction to normal forms of the pitchfork and Hopf
bifurcations has been carried out in
Refs. \citep{Giannakopoulos99,Redmond02}.  If we cast the equation
for the mean field dynamics of our model in this form, we can use the
results in Refs. \citep{Giannakopoulos99,Redmond02} to determine the
functional dependence of the normal form coefficients on the
parameters $D$, $\varepsilon$, and $\tau$. This can be achieved by a
series expansion of Eq.~(\ref{XXuniform}) up to the third order and
rescaling of time.

\begin{figure}
\includegraphics[angle=-90,width=6cm]{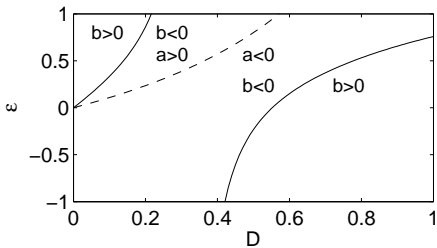}
\caption{\label{manifoldp} Stability diagram for the trivial
equilibrium resulting from the analysis of the pitchfork bifurcation.
The solid and the dashed lines respectively depict the b=0 and a=0
contour lines. The stability diagram for the pitchfork bifurcation is
time delay independent.}
\end{figure}

The normal form of the pitchfork bifurcation reads
\begin{equation}
\dot{z}=az+bz^3, 
\end{equation}
where $z$ is a coordinate on the center manifold. 
The normal form coefficients are
\begin{eqnarray}
\label{pita}
a & = &\frac{\varepsilon-\varepsilon_{\rm st}}
{\varepsilon_{\rm st}(1-\tau_0)}\\
\label{pitb}
b&=&\frac{B_1-12 D B_2}{384 (1-\tau_0) D^3},
\end{eqnarray}
where $B_1=\varepsilon^3(81 D^3+108 D^2 +144 D-64)$,
$B_2=\varepsilon^2(9D^2+24D-16)$ and
$\tau_0=-\sqrt{2}\exp(-1/4D)\tau/\pi$. Setting $a=0$ and solving
Eq.~(\ref{pita}) for $\varepsilon$ we again find the critical coupling
of Eq.~(\ref{ec}). One can show that $b<0$ for $D>0$ and
$\varepsilon=\varepsilon_{\rm st}$ (i.e. $a=0$). Consequently, the
pitchfork bifurcation at $\varepsilon_{\rm st}$ is always
supercritical. The stability diagram resulting from center manifold
reduction for the pitchfork bifurcation is shown in
Fig.~\ref{manifoldp}.

The normal form of the Hopf bifurcation in polar coordinates $r$ and
$\theta$ on the center manifold reads
\begin{equation}
\label{normhopf}
\dot{r}=\mu r +\alpha r^3\ ,\;\;
\dot{\theta}= \omega + \rho r^2.
\end{equation}
The coefficients determining the stability of the trivial equilibrium
and the order of the Hopf bifurcation are $\mu$ and $\alpha$
\citep{Strogatz}. Expressed in system parameters, they read
\begin{eqnarray}
\label{hopfmu}
\mu&=&\left(\frac{\varepsilon}{\varepsilon_{\rm st}}-\frac{1}{\cos\varphi}\right)\left(\frac{1-\tau_0}{(1-\tau_0)^2+\varphi^2}-\frac{\sin\varphi}{2}\right),\\
\label{hopfalpha}
\alpha&=&\frac{B_1 B_3-B_2(1-3\tau_0+2\cos^2\varphi)}{128([1-\tau_0]^2+\varphi^2)D^2},
\end{eqnarray}
where $B_3=(\cos^2\varphi-\tau_0)/(4 D \cos\varphi)$, $\varphi=\omega_0\tau_0$ and
$\omega_0=\tan\varphi$.

\begin{figure}
\includegraphics[angle=-90,width=6cm]{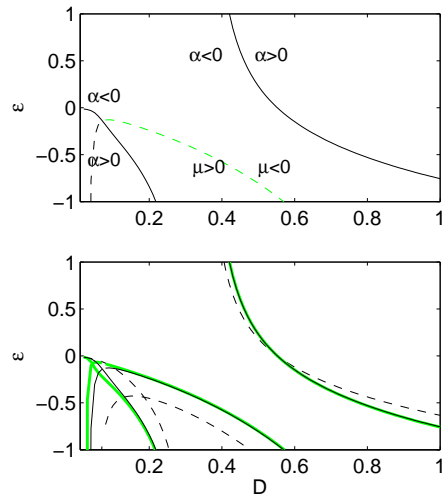}
\caption{\label{manifoldhopf} Primary solutions of  
$\mu=0$ (Eq. \ref{hopfmu}) and $\alpha=0$ (Eq. \ref{hopfalpha}) in the
$(\varepsilon,D)$-parameter-space. Upper panel: The boundaries $\mu=0$
(dashed line) and $\alpha=0$ (solid line) for a system with
$\tau=100$. The black dashed line and the gray dashed line depicts the
parameter values of the subcritical and supercritical Hopf
bifurcation, respectively. The lower panel shows the same curves for a
system with $\tau=10$ (dashed line), $\tau=100$ (black solid line) and
$\tau=1000$ (gray solid line).}
\end{figure}

Setting $\mu=0$ and solving Eq.~(\ref{hopfmu}) for $\varepsilon$
yields the critical coupling as a function of the noise strength
$\varepsilon_{\rm osc}(D)$ which coincides with Eq.~(\ref{omegaeps2}).
Setting the first Lyapunov coefficient $\alpha=0$, we can find
$\varepsilon_{\alpha=0}(D)$. The two functions $\varepsilon_{\rm
osc}(D)$ and $\varepsilon_{\alpha=0}(D)$ intersect at a noise level
$D_{\rm H}$ denoting the parameter values for which the Hopf
bifurcation changes from supercritical to subcritical.
The stability diagram resulting from the analysis
of the Hopf bifurcation is shown in Fig.~\ref{manifoldhopf}.

Let us now discuss the bifurcation properties in the limit of large
and small time delays as well as vanishing noise and compare them with
those of a single oscillator system. The critical coupling
$\varepsilon_{\rm st}$ of the pitchfork bifurcation is time delay
independent and goes to zero for vanishing noise. However, the
critical coupling of the Hopf bifurcation depends on the time delay
(see lower panel in Fig. \ref{manifoldhopf}). As the time delay
increases, the maximum of the primary Hopf bifurcation line
$\varepsilon_{\rm osc-}^1$ approaches the origin in the $(\varepsilon,
D)$ plane meaning that oscillations may occur at an arbitrary small
feedback strength for the properly tuned noise level. This should be
contrasted to the dynamics of a single noise-free oscillator with
time-delayed feedback that only exhibits oscillations at strong
negative feedback $(\epsilon<-1)$.  For very small time delays
$\tau\to0$, the critical coupling strength $\varepsilon^{l,k}_{\rm
osc\mp}\to\mp\infty$.

\subsection{Coherence resonance and system size effects}

\label{coherence1}
The system studied in this paper exhibits the phenomenon of
coherence resonance \citep[e.g.][]{Kiss03,Miyakawa02,Pikovsky97,Sigeti89} 
and array-enhanced resonance \citep{Pikovsky02,Zhou01}. 

Let us discuss this in turn. If our system adopts an oscillatory
state, the double-well potentials of the elements are tilted
asymmetrically, due to their coupling to the delayed mean field; that
is, the potential barriers separating the two wells are periodically
rising the lowering. If the period of this oscillation $T$ matches the
time scale $\tau_{\rm K}$ of the noise-induced inter-well fluctuation,
i.e., if the mean field oscillations synchronize with the hopping rate,
we can expect that the number of elements contributing to the
oscillation and consequently the order of the oscillatory state
reach a maximum. In this spirit the time scale matching condition
for such a synchronization, which is given through
\begin{equation} 
\label{match}
2\tau_{\rm K}=T,
\end{equation} 
is a reasonable condition for the maximum order of the oscillatory
state \citep{Gammaitoni98}.

\begin{figure}
\includegraphics[angle=-90,width=\hsize]{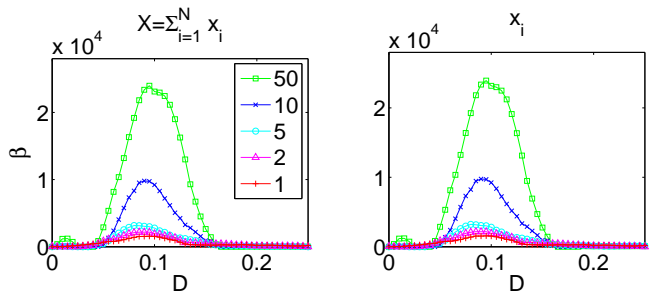}
\caption{\label{coherence} The coherence of the oscillatory
states $\beta$ as a function of the noise strength $D$ for systems of
different size $N$. The time delay and the coupling strength are
$\tau=100$ and $\varepsilon=-0.2$, respectively.  Left panel: The
coherence of the mean field oscillations. Right panel: The coherence
of a single element $x_i$ out of the $N$ network elements.}
\end{figure}

To quantify the order (i.e. coherence) of the oscillatory state we
introduce the coherence parameter $\beta=H \omega_{\rm peak}/\Delta\omega$, 
where H is the height
of the main spectral peak at $\omega_{\rm peak}$ and $\Delta\omega$ is its
halfwidth. Using the Langevin model (\ref{languniform}), the coherence
measure $\beta$ is determined as a function of the noise strength 
and in  Fig.~\ref{coherence} compared for systems of different size $N$. 

\begin{figure}
\includegraphics[angle=-90,width=4cm]{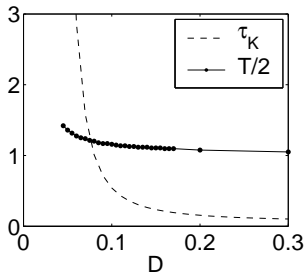}
\caption{\label{reson} The Kramers time $\tau_{\rm K}$ and the half
the period of the mean field oscillations (in units of the time delay)
as a function of the noise strength. The parameters are $\tau=100$ and
$\varepsilon=-0.2$.}
\end{figure}

Clearly, the coherence curves have a maximum. The noise strength
maximizing the coherence is $D_{\rm S}\approx0.08$.  This noise
strength can also be derived from the time scale matching condition in
Eq.~(\ref{match}). The Kramers time $\tau_{\rm K}=1/p$ is given
through Eq.~(\ref{hopuniform}) and the period of the oscillations $T$
beyond the critical coupling can be determined numerically. In
Fig.~\ref{reson} the two time scales are plotted as function of the
noise strength.  The curves intersect at D=0.08 substantiating the
consistency of theory and Langevin model.

\begin{figure}
\includegraphics[angle=-90,width=\hsize]{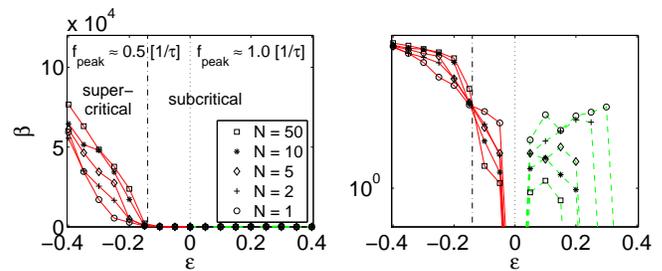}
\caption{\label{varn} The coherence $\beta$ as function
of the coupling strength. For $\varepsilon<0$ and $\varepsilon>0$ the
spectral peak frequency is $f_{\rm peak}=\omega_{\rm
peak}/2\pi\approx0.5\;1/\tau$ and $f_{\rm peak}\approx1.0\;1/\tau$,
respectively. The dash-dotted vertical line depicts
$\varepsilon^1_{\rm osc-}$ and consequently separates domain of the
macroscopic (i.e. supercritical) mean field oscillations from the
domain of subcritical coherence. The right panel shows the same as the
left, but has a logarithmic scale for $\beta$, which helps to
uncover the weak subcritical coherence properties.}
\end{figure}

The resonance curves in Fig.~\ref{coherence} show that the coherence
of the oscillatory states increases with increasing $N$, a property
which was reported for other systems and is sometimes referred to as
array-enhanced resonance \citep{Zhou01}. Interestingly, the
enhancement of the temporal regularity with increasing system size is
only observed for macroscopic mean field oscillations, while the
inverse holds for ``subcritical coherence''. That is, the coherence
observed in the power spectra of subcritical mean field fluctuations
(i.e., for $|\epsilon|<|\epsilon_{\rm osc\pm}|)$ decays inversely
proportional to the number of elements in the network, and becomes
negligible for $N>10$. This is show in Fig.~\ref{varn}. Qualitatively,
the same dependency on the system size is found if the delayed average
does not include the delayed element itself, i.e., the element $x_i$
is coupled to $X_i(t-\tau)=\sum^{N-1}_{{j=1},{j\neq i}}x_j$.

\section{Two delays}

\label{twodelays}

\subsection{Langevin model}

We want to generalize the above system by introducing multiple time
delays and non-uniform coupling terms. Let us carry out the
generalization progressively and study first the dynamics of a
bistable element network with a discrete bimodal delay distribution
(i.e. with two time delays) and uniform coupling. Assuming that the
time delay of the interaction between two elements is entirely
determined by the ``transmitting'' element the system dynamics is
described by the following set of Langevin equations,
\begin{equation}
\dot{x_i}=x_i-x_i^3+\frac{\epsilon}{2}X_1(t-\tau_1)
                       +\frac{\epsilon}{2}X_2(t-\tau_2)+\sqrt{2D}\xi(t)
\label{lang2}
\end{equation} 
where
\begin{equation}
X_1(t)=\frac{2}{N}\sum_{j=1}^{N/2} x_j(t) 
\end{equation}
and
\begin{equation}
X_2(t)=\frac{2}{N}\sum_{j=N/2+1}^N x_j(t),
\end{equation}
are the mean fields of the elements associated with time delay $\tau_1$
and $\tau_2$, respectively. Here, it is assumed that the number of oscillators
is the same in both group.

\subsection{Dichotomous theory}

We want to use the dichotomous theory in order to study the mean field
dynamics of model (\ref{lang2}). Thus, the theory developed in
Sect.~\ref{dimod1} has to be extended accordingly. In order to describe the
collective dynamics of the two-delay system, two equations are needed,
respectively describing the mean field evolution of the oscillator
group associated with $\tau_1$ and $\tau_2$,
\begin{equation}
\label{dicho2}
\dot X_{1,2}(t)=p_{\rm 12}-p_{\rm 21}-(p_{\rm 12}+p_{\rm 21})X_{1,2}(t).
\end{equation}
The mean field of the entire system then is, $X=(X_1+X_2)/2$,
and  the hopping probabilities are given by,
\begin{equation}
p_{\rm 12,21}=\frac{\sqrt{2\mp 3\alpha_2}}{2\pi}
\exp\left(-\frac{1\mp4\alpha_2}{4 D}\right),
\end{equation}
where $\alpha_2=\varepsilon\left [X_1(t-\tau_1)+X_2(t-\tau_2)\right
]/2$.  As for the model with a single (i.e. uniform) time delay, the
numerical integration of the Langevin system (\ref{lang2}) reveals
pitchfork and Hopf bifurcations describing the transitions to
nontrivial stationary states and oscillatory states, respectively.
The critical couplings for the bifurcations can be found
with a linear stability analysis of Eqs.~(\ref{dicho2}) near the
trivial state $X=0$. The procedure, which is analogous to the stability
analysis carried out in Sect. (\ref{dimod1}), yields the transcendent
equation for the complex eigenvalue $\lambda$,
\begin{equation}
\label{lambda2}
\lambda=\frac{S\pm \sqrt{S^2-4\Delta}}{2}.
\end{equation}
Here, $S$ and $\Delta$ respectively are the trace and the determinant
of the Jacobian matrix
\begin{equation}
\label{jacob1}
J=c
\begin{pmatrix}
g_1+d & g_2 \\
g_1 & g_2+d
\end{pmatrix},
\end{equation}
where the matrix elements are given through 
$c=-\sqrt{2}\exp(-1/4D)/8\pi D$, 
$g_{1,2}=\varepsilon(3D-4)\exp(-\lambda\tau_{1,2})$ and $d=8D$.
For a positive coupling Eq.~(\ref{lambda2}) has always a real 
eigenvalue. At a certain critical coupling
\begin{equation}
\label{pcrit2d}
\varepsilon_{\rm st}=\frac{4D}{4-3D}
\end{equation}
the eigenvalue becomes positive indicating the pitchfork bifurcation. 
This bifurcation is time delay independent and is thus identical with 
those found for the system with uniform time delays [cf. Eq.~(\ref{ec})].

For finite $\bar{\tau}=(\tau_1 + \tau_2)/2$ and $\varepsilon$,
Eq.~(\ref{lambda2}) possesses also a finite number of complex solutions.
The critical couplings of the corresponding unstable modes (i.e. of the
Hopf bifurcation) are given by the following set of equations,
\begin{eqnarray}
\label{omegaeps2a} 
\omega\bar{\tau} &=& -\frac{\sqrt{2}}{\pi}
\exp(-1/4D)\bar{\tau}\tan\omega\bar{\tau},\\
 \label{omegaeps2b} 
\varepsilon_{\rm osc}&=& \frac{8 D\pi\omega}
{(3D-4)(\sqrt{2}\exp[-1/4D] J_{\rm s}-\pi\omega J_{\rm c})},
\end{eqnarray} 
Here 
\begin{eqnarray}
\label{eqjs}
J_{\rm s} &=& \frac{1}{2}(\sin\omega\tau_1+\sin\omega\tau_2)
\;=\;\sin\omega\bar{\tau}\cos\omega\sigma\\
J_{\rm c} &=& \frac{1}{2}(\cos\omega\tau_1+\cos\omega\tau_2)
\;=\;\cos\omega\bar{\tau}\cos\omega\sigma,
\label{eqjc}
\end{eqnarray}
where $\sigma=|\tau_1-\tau_2|/2$. The above set of equations
for the critical coupling is the two-delay analog to
Eq.~(\ref{omegaeps1}) and (\ref{omegaeps2}).  Again, we find a
multiplicity of solutions, leading to the multi-stability of the
system in a certain area of the parameter space. Furthermore, Eqs. 
(\ref{omegaeps2a}),(\ref{omegaeps2b}
show that while the frequencies of the oscillatory states only depend
on the mean time delay $\bar{\tau}$, the critical coupling strengths
of the Hopf bifurcations depend additionally on $\sigma$.

\subsection{Phase diagrams}

\begin{figure}
\includegraphics[width=8cm]{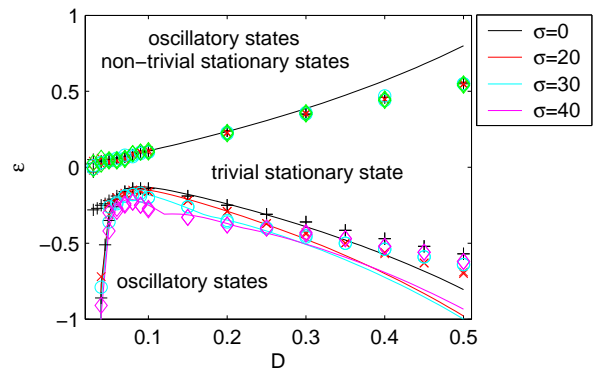}
\caption{\label{comp2d} Phase diagram of the globally
coupled two-delay network determined using the
dichotomous model (solid lines) and numerical simulations
of the Langevin model (markers). The phase diagram is show
for different $\sigma=|\tau_1-\tau_2|/2$. The mean
time delay is $\bar{\tau}=100$.}
\end{figure}

\begin{figure*}
\includegraphics[angle=-90,width=13cm]{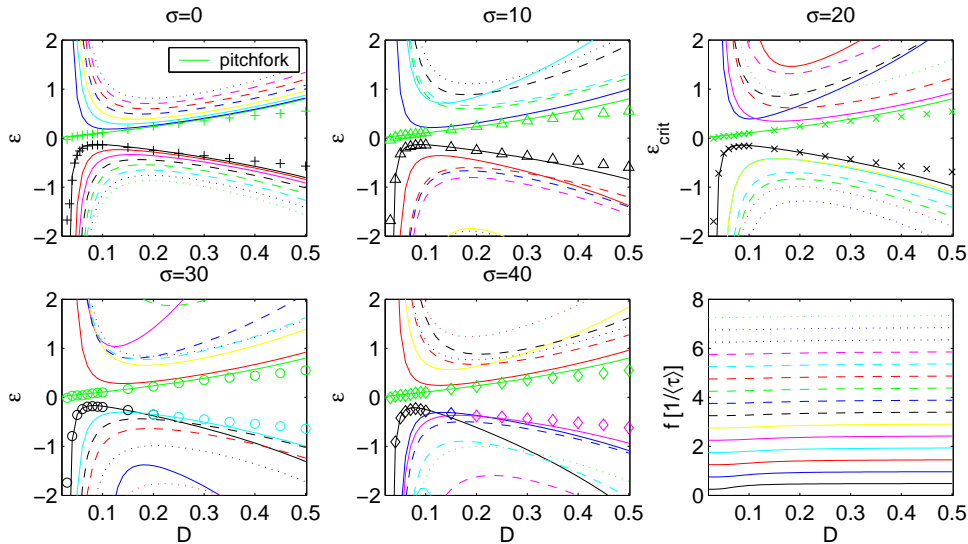}
\caption{\label{multi2d} Upper panel and two lower left panels:
Phase diagrams of our globally coupled two-delay network with
$\bar{\tau}=100$ and $\sigma=0,\;10,\; 20,\;30,\;40$. The green line depicts
the critical coupling of the pitchfork bifurcation and the other lines
depict those of the primary Hopf bifurcation as well as some higher
order solutions (i.e. solutions 1-15) of Eq. \ref{omegaeps2a} and
\ref{omegaeps2b}. The markers depict the first bifurcation at 
$\varepsilon<0$ and $\varepsilon>0$ resulting from numerical
simulations of the Langevin model (in these simulations, starting with
$\varepsilon=0$, the coupling strength is increased until a
bifurcation occurs). Matching colors of markers and lines mean that
the bifurcation type and associated frequency are in agreement. Lower
right panel: The frequencies of the corresponding unstable modes. They
do not depend on $\sigma$, but slightly vary with the noise strength
$D$.}
\end{figure*}

The phase diagram and the frequencies of the unstable oscillatory
modes of the two-delay system are theoretically determined using
Eqs.~(\ref{pcrit2d})-(\ref{omegaeps2b}) and compared with numerical
findings resulting from simulations of the Langevin model
(\ref{lang2}). The phase diagram is shown in Fig.~\ref{comp2d} for
different $\sigma$. The phase diagrams including higher order solutions
of Eq.~(\ref{omegaeps2a}) and (\ref{omegaeps2b}) are presented in
Fig.~\ref{multi2d}. Also, the frequencies of the corresponding
unstable modes (which are $\sigma$-independent) are shown in this
Figure. The Figures show that near the bifurcation points the
predictions by the dichotomous theory are reasonably good for weak
noise in the range $(0.07 \lesssim D\lesssim 0.3)$.

Furthermore, we find that the first bifurcation of the
trivial equilibrium at $\varepsilon>0$ is always a pitchfork
bifurcation.  The first bifurcation at $\varepsilon<0$ is a Hopf
bifurcation, which for $\sigma<30$ is determined by the primary
solution of Eq.~(\ref{omegaeps2a}) and (\ref{omegaeps2b}), while for
$\sigma>30$, depending on the noise intensity, the first transition
may also be determined by higher order solutions associated with
higher frequencies.


\section{Multiple delays}

\label{multidelays}

\subsection{Langevin model}


In this Section we further generalize our delayed feedback system by
introducing multiple time delays and study the stability properties in
dependence of the statistical moments of an arbitrary time delay
distribution.

The general Langevin model with many time delays reads
\begin{equation}
\label{manyij}
\dot{x}_i=x_i-x^3_i+\frac{\varepsilon}{N}\sum_{j=1}^N 
x_j(t-\tau_{ij})+\sqrt{2D}\xi(t).
\end{equation} 

Such general models in which the time delays depend on both the
``transmitting'' and the ``receiving'' element cannot directly be described in
terms of a mean field theory. However, the system becomes
mathematically tractable if we assume that the time delays only
depend on the transmitting elements $j$,
\begin{equation}
\label{manyj}
\dot{x}_i=x_i-x^3_i+\frac{\varepsilon}{N}\sum\limits_{j=1}^N 
x_j(t-\tau_{j})+\sqrt{2D}\xi(t).
\end{equation}

\begin{figure}
\centerline{\includegraphics[width=6cm]{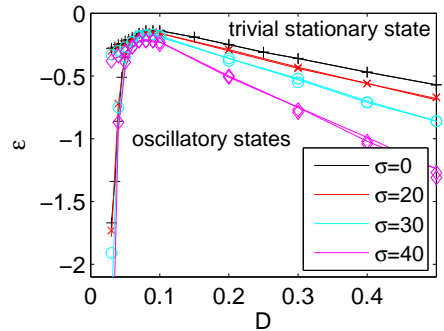}}
\caption{\label{compmod} The critical coupling of the Hopf bifurcation
as a function of the noise strength $D$ for different $\sigma$
of the Gaussian time delay distribution with $\bar{\tau=100}$. The
markers and the solid lines depict the critical couplings resulting
from model (\ref{manyij}) and model (\ref{manyj}), respectively.}
\end{figure}

In order to check if such a simplification is justified, numerical
simulations of model (\ref{manyij}) and (\ref{manyj}) are carried out
and compared. In these simulations the distribution of the time delays
is Gaussian, i.e., it is fully determined by its mean $\bar{\tau}$ and
standard deviation $\sigma$. Fig.~\ref{compmod}, comparing the
critical coupling strength of the Hopf bifurcation for different
$\sigma$, suggests that the above simplification is justified in order
to study the stability properties of a bistable-element-network with
time delays.

This surprising result not only renders possible an analytical
description of networks with distributed delays but also implies
that the number of operations, which have to be carried out to study
such systems numerically, can be reduced from ${\cal O}(N^2)$ to
${\cal O}(N)$. 

\subsection{Dichotomous theory}

Let us now develop the dichotomous theory for the globally coupled
bistable-element-network with distributed delays.

For that purpose we coarse grain system (\ref{manyj}). 
The coarse graining
is accomplished as follows: The range of possible time delays is
divided up in $M$ intervals $I_k\;\{k=1,2,\ldots,M\}$. The size of the
intervals $\Delta_k$ is chosen, so that the number of bistable
oscillators associated with a delay, fitting in a particular interval,
is for each interval the same $m=N/M$. In this way oscillator groups
are formed whose mean field can be expressed as,
\begin{equation}
\Omega_k(t)\equiv\frac{1}{m}
\sum_{\tau_j\in I_k} x_j(t),
\end{equation}
where $I_k\equiv[\tau_k,\tau_{k+1}[$, $\tau_k=\sum_{l=1}^{k-1}\Delta_l$
and $j=1\ldots N$.

Assuming that $\Delta_k\ll\bar{\tau}/\sigma$, where $\bar{\tau}$ and
$\sigma$ are the mean and the standard deviation of the time delay
distribution, Eq.~(\ref{manyj}) can then be approximated by,
\begin{equation}
\label{omegadyn}
x_i=x_i-x_i^3+\frac{\varepsilon}{M}\sum\limits_{k=1}^M \Omega_k(t-\tau_k)
+\sqrt{2D}\xi(t).
\end{equation}

The master equations expressing the dynamics of system (\ref{omegadyn})
in terms of occupation probabilities read,
\begin{eqnarray}
\label{ndot1}
\dot{n}_{1,k}  &=&  -p_{12}n_{1,k}+p_{21}n_{2,k}\\
\dot{n}_{2,k}  &=&  p_{12}n_{1,k}-p_{21}n_{2,k}.
\end{eqnarray} 
Here the hopping probabilities are given by,
\begin{equation}
p_{12,21}=\frac{\sqrt{2\mp 3\alpha_3}}{2\pi}
\exp\left(-\frac{1\mp4\alpha_3}{4 D}\right),
\end{equation} 
where $\alpha_3=(\varepsilon/M)\sum_{k=1}^M \Omega_k (t-\tau_k)$.

For large oscillator groups $(m\to\infty)$,
$\Omega_k=n_{1,k}s_1+n_{2,k}s_2=n_{2,k}-n_{1,k}$ holds. With this and
the probability conservation $n_{1,k}+n_{2,k}=1$ we can find the
following set of equations:
\begin{equation}
\label{manyhop}
\dot \Omega_k(t) = p_{12}-p_{21}-(p_{21}+p_{12})\Omega_k(t).
\end{equation}
The Jacobian matrix of this system is given through,
\begin{equation}
J=c
\begin{pmatrix}
g_1+d & g_2 & \dots & g_M \\
g_1 & g_2+d & \dots & g_M \\
\hdotsfor{4}              \\
g_1 & g_2 & \dots & g_M+d \\
\end{pmatrix},
\end{equation}  
where $c=-\sqrt{2}\exp(-1/4D)/(4M\pi D)$, 
$g_k=\varepsilon(3D-4)\exp(-\lambda\tau_k)$ and $d=4MD$.
With this Jacobian the characteristic equation, determining
the stability of the trivial equilibrium $X=0$, becomes
\begin{equation}
\label{ceq1}
(dc-\lambda)^{M-1}\left(c\left[d+\sum_{k=1}^M g_k\right]-\lambda\right)=0.
\end{equation}

Setting $\lambda=0$ and solving Eq.~(\ref{ceq1})
for $\varepsilon$ yields the critical coupling for the pitchfork
instability,
\begin{equation}
\label{epspf}
\varepsilon_{\rm st}=\frac{4D}{4-3D}.
\end{equation}
It is time delay independent and thus identical with that
found in previous Sections of this paper.

The properties of the Hopf bifurcation (i.e. the frequencies of the
unstable modes $\omega$ and the critical coupling $\varepsilon_{\rm
osc}$ can be found by substituting $\lambda=\mu+{\rm i}\omega$ into
Eq.~(\ref{ceq1}), separating real and imaginary parts and setting
$\mu=0$. This yields,
\begin{equation} 
\label{manyomega}
\omega\bar{\tau} = -\frac{\sqrt{2}}{\pi}
\exp(-1/4D)\bar{\tau}\frac{I_{\rm s}}{I_{\rm c}},
\end{equation}
where 
\begin{equation}
I_{\rm s}=\frac{1}{M}\sum_{k=1}^{M}\sin\omega\tau_k,\;\:
I_{\rm c}=\frac{1}{M}\sum_{k=1}^{M}\cos\omega\tau_k.  
\end{equation}
For large systems $N\to\infty$, the number of groups $M\to\infty$
and thus
\begin{equation}
\label{integrals}
I_s=\int\limits_0^{\infty}P(\tau)\sin\omega\tau d\tau,\;\:
I_c=\int\limits_0^{\infty}P(\tau)\cos\omega\tau d\tau,
\end{equation} 
where $P(\tau)$ is the time delay distribution function.

We can express the time delay distribution function in terms of
cumulants $K_n$ \citep{Kampen03,Risken89} and solve the integrals in
(\ref{integrals}):
\begin{equation}
I_s=\sin(g_1)\exp({g_2}),\;\:
I_c=\cos(g_1)\exp({g_2}),
\end{equation}
where
\begin{eqnarray}
g_1 & = & \sum_{m=0}^\infty \frac{({\rm i}\omega)^{2m+1}}
{{\rm i}(2m+1)!}K_{2m+1},\\ 
g_2 & = & \sum_{m=1}^\infty\frac{({\rm i}\omega)^{2m}}{(2m)!}K_{2m}.
\end{eqnarray}
Consequently, 
\begin{equation}
\label{tang1}
\frac{I_{\rm s}}{I_{\rm c}}=\tan(g_1).
\end{equation}
Since for symmetric distribution functions all odd cumulant moments
except the first one $K_1=\bar{\tau}$ are zero, $I_{\rm s}/I_{\rm
c}=\tan\omega\bar{\tau}$ holds. That is, in the case of a symmetric
distribution of the time delays, the frequencies of the unstable modes
in Eq.~(\ref{manyomega}) depend only on the mean time delay.

Let us now determine the critical coupling of the Hopf bifurcation.
For large time delays $\bar{\tau}\gg\tau_K$ the low-order solutions of
the transcendental equation (\ref{manyomega}) yield frequencies
$\omega\ll 1$. Thus the real part of equation (\ref{ceq1}) can be
linearized near $\omega=0$ and the critical coupling of the Hopf
bifurcation becomes,
\begin{equation} 
\varepsilon_{\rm osc}= \frac{4 D\pi\omega}
{(3D-4)\left(\frac{1}{N}\sqrt{2}\exp(-1/4D)I_{\rm s}
-\left[1-\frac{1}{N}\right]\pi\omega I_{\rm c}\right)}.
\end{equation} 
Then, for large systems $N\to\infty$ the critical coupling is,
\begin{equation}
\label{epsosc}
\varepsilon_{\rm osc}= \frac{4 D}
{(4-3D)I_{\rm c}},
\end{equation}
with $I_{\rm c}=3\sin(\omega\bar{\tau})\sin(5\omega\sigma/3)/
(5\omega\sigma)$ and $I_{\rm c}
=\cos(\omega\bar{\tau})\exp(-\omega^2\sigma^2/2)$ for 
uniform and Gaussian distributions, respectively.


\subsection{Phase diagrams}

\begin{figure}
\centerline{\includegraphics[width=8cm]{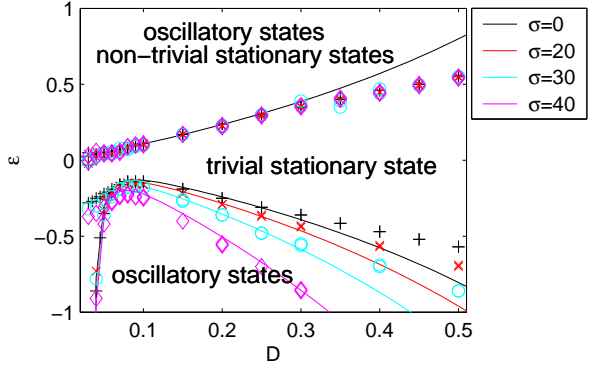}}
\caption{\label{compmulti} Phase diagram of the globally coupled
bistable-element-network with uniformly distributed time delays
derived from the theoretical model (solid lines) and numerical
simulations of the Langevin model (markers). The phase diagram is
shown for different standard deviations $\sigma$ of the delay
distribution function.  The mean time delay is $\bar{\tau}=100$.}
\end{figure}
\begin{figure}
\centerline{\includegraphics[width=5cm]{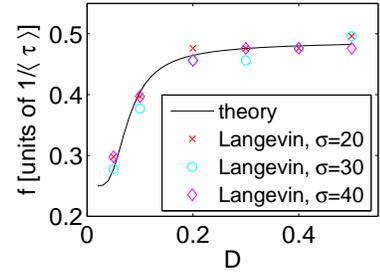}}
\caption{\label{freqmulti} The frequencies of the unstable modes at
the bifurcation points resulting from the Langevin model (markers)
and the the dichotomous model (solid line), which are
(see Eq.~(\ref{manyomega}) independent of $\sigma$. For uniform and
Gaussian distributions the frequencies depend only on the mean time
delay (see Eq. \ref{manyomega} and \ref{tang1}).}
\end{figure}

Eqs.~(\ref{epspf}), (\ref{manyomega}) and (\ref{epsosc}) are used to
determine the phase diagram and the frequencies of the unstable
oscillatory modes $f=\omega/(2\pi)$ of a bistable element network with
uniformly distributed time delays \footnote{This should not be
confused with uniform time delays, which means that the delay for each
coupling is the same.}. The theoretical predictions are compared with
numerical simulations of the Langevin model (\ref{manyij}). The
number of bistable elements in these simulations is $N=300$. The results are shown
in Figs.~\ref{compmulti} and \ref{freqmulti}.

Again, we find that near the transition points and for weak noise
intensities the predictions of the dichotomous theory are reasonably
good. Consequently, the Langevin models (\ref{manyij}) and
(\ref{manyj}) are in this regime equivalent as regards the dynamical
properties of the mean field.


\begin{figure}
\centerline{\includegraphics[width=\hsize]{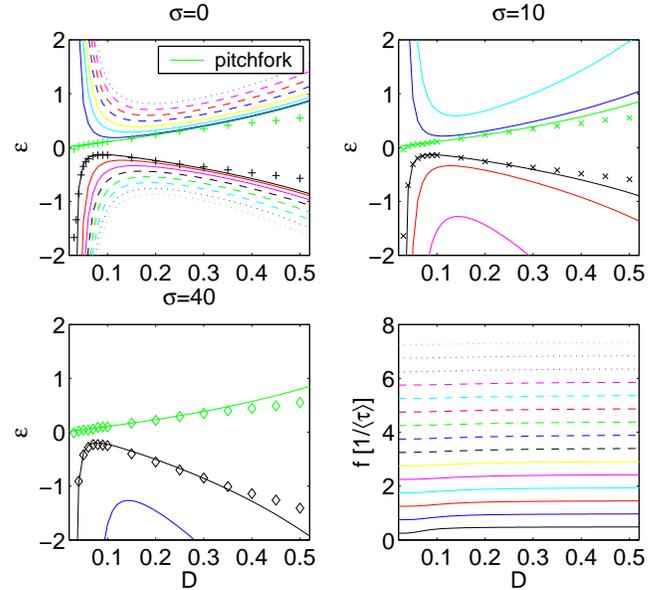}}
\caption{\label{phasemulti} Same as in Fig.~\ref{multi2d} but this time
for  networks with uniformly distributed time delays with $\tau=100$ and
$\sigma=0,\:10,\:40$} 
\end{figure}

Eqs.~(\ref{manyomega}) and (\ref{epsosc}) have a multiplicity of
solutions meaning that multistability is also present in our system in
the limit of continuously distributed delays. The bifurcation diagrams
including the higher order solutions are shown in Fig.~\ref{phasemulti}. The
Figure shows that unlike the two-delay system, the first
transition at $\varepsilon<0$ is always determined by the primary
solution associated with the frequency $f\approx0.5\;\tau$. 

The comparison of the phase diagrams for delay distribution functions
of different widths $\sigma$ shows that the regions of oscillatory
states in the parameter space are reduced with increasing $\sigma$.
This trend was already apparent in the two-delay system, although less
pronounced. These findings suggest that nonuniformity of the time
delays inhibits the occurrence of Hopf bifurcations and consequently
increases the stability of the trivial equilibrium.

Eventually, we like to mention that the coherence resonance phenomenon
discussed in Sect.~\ref{coherence1} is also present in systems with
multiple delays in the oscillatory domain of the phase diagram.

\section{Nonuniform coupling}

\label{twocouplings}

\subsection{Langevin model} 

The collective dynamics of the bistable-element-networks described
above, is restricted to periodic oscillations and stationary
states. In this Section we want to check whether the complexity of the
dynamics is increased if instead of the uniform coupling, non-uniform
couplings are applied. To this end, we extend the two time delay model
(\ref{lang2}) by introducing two different coupling strengths. The
Langevin equations of the new model read,
\begin{equation}
\begin{aligned}
\label{lang2c}
\dot{x_i} & = x_i-x_i^3+\frac{\epsilon_1}{2}X_1(t-\tau_1)
            +\frac{\epsilon_2}{2}X_2(t-\tau_2)+\sqrt{2D}\xi(t)\\
\dot{x_j} & =  x_j-x_j^3+\frac{\epsilon_2}{2}X_1(t-\tau_1)
            +\frac{\epsilon_1}{2}X_2(t-\tau_2)+\sqrt{2D}\xi(t),
\end{aligned}
\end{equation}
where $i=1,\ldots,N/2$ and $j=N/2+1,\ldots,N$. The elements $x_i$ and
$x_j$ belong to a group of bistable oscillators which are associated
with time delay $\tau_1$ and $\tau_2$, respectively. It is assumed
that the two groups are of equal size. The above set of equations
describes a system in which each element couples to all the elements
belonging to the same group with a coupling strength $\varepsilon_1$
and to all the elements of the other group with $\varepsilon_2$; that
is, the two coupling parameters indicate the strength of the
intra-group coupling $(\varepsilon_1)$ and inter-group coupling
$(\varepsilon_2)$, respectively.

\subsection{Dichotomous model}

\label{sectwocoup}
We apply the dichotomous theory to system (\ref{lang2c}) and proceed
in a manner analogous to the previous sections.

The evolution of the mean field of each group of oscillators is
described by
\begin{equation}
\label{mf2c}
\dot X_{1,2}(t)=p^{1,2}_{12}-p^{1,2}_{21}-
               (p^{1,2}_{12}+p^{1,2}_{21})X_{1,2}(t).
\end{equation}
Here, the hopping probabilities are,
\begin{eqnarray}
p^1_{12,21}&=&\frac{\sqrt{2\mp 3\alpha_4}}{2\pi}
\exp\left(-\frac{1\mp4\alpha_4}{4 D}\right),\\
p^2_{12,21}&=&\frac{\sqrt{2\mp 3\alpha_5}}{2\pi}
\exp\left(-\frac{1\mp4\alpha_5}{4 D}\right),
\end{eqnarray}
where 
\begin{equation}
\alpha_{4,5}=\left [\varepsilon_1 X_{1,2}(t-\tau_1)
         +\varepsilon_2 X_{2,1}(t-\tau_2)\right]/2.
\end{equation} 

Next a linear stability analysis is carried out. The linearization
of Eq. \ref{mf2c} about the trivial equilibrium yields the Jacobian,
\begin{equation}
\label{jnuc}
J=-c
\begin{pmatrix}
[3D-4]\varepsilon_1{\rm e}^{-\lambda\tau_1}+d & [3D-4]\varepsilon_2{\rm e}^{-\lambda\tau_2}, \\
[3D-4]\varepsilon_2{\rm e}^{-\lambda\tau_1} & [3D-4]\varepsilon_1{\rm e}^{-\lambda\tau_2}+d
\end{pmatrix},
\end{equation}
where $c$ and $d$ are the same as in Eq.~(\ref{jacob1}).

Substituting the trace $S$ and determinant $\Delta$ of the Jacobian
matrix (\ref{jnuc}) into the characteristic equation
\begin{equation}
\label{lambda4}
\lambda=\frac{S\pm \sqrt{S^2-4\Delta}}{2},
\end{equation}
and keeping the intra-group coupling strength $\varepsilon_1$ fixed,
yields the critical coupling for the pitchfork bifurcation, which can
occur for positive and negative inter-group feedbacks,
\begin{equation}
\label{epsinterst}
\varepsilon^{\rm st}_2=\pm\left(\varepsilon_1+\frac{8D}{3D-4}\right).
\end{equation}
In order to find the critical values for the Hopf bifurcation
$\varepsilon^{\rm osc}_2$, we substitute $\lambda=\mu+{\rm i}\omega$
into the characteristic equation and set $\mu=0$. Then, the separation
of real and imaginary part yields the two equations, $f_{\rm
r}(\omega,\varepsilon_2)=0$ and $f_{\rm i}(\omega,\varepsilon_2)=0$,
where
\begin{eqnarray}
\label{rootsof2ca}
f_{\rm r}(\omega,\varepsilon_2) &=&  E_1\varepsilon_1 J_s\omega
+\frac{E_1^2}{4}(\varepsilon^2_1
-\varepsilon^2_2)\cos(2\omega\bar{\tau})\nonumber\\
&+& E_2\varepsilon_1 J_c+\frac{8}{\pi^2}\exp(-1/2D)-4\omega^2,\\
f_{\rm i}(\omega,\varepsilon_2) &=&  E_1\varepsilon_1 J_c\omega
+\frac{E_1^2}{4}(\varepsilon^2_1
-\varepsilon^2_2)\sin(2\omega\bar{\tau})\nonumber\\
&+& E_2\varepsilon_1 J_s+\frac{8\sqrt{2}\omega}{\pi}\exp(-1/4D).
\label{rootsof2cb}
\end{eqnarray}
Here, $E_1=\sqrt{2}(3D-4)\exp(-1/4D)/(\pi D)$ and
$E_2=E_1\sqrt{2}\exp(-1/4D)/\pi$. The terms $J_s$ and $J_c$ are given
by Eq.~(\ref{eqjs}) and (\ref{eqjc}), respectively.

For finite $\bar{\tau}$ and $\varepsilon_2$ the above set of equations
has a finite number of roots $(\varepsilon^{\rm osc}_2,\omega)$, which
can be found numerically.

\subsection{Phase diagram}

\begin{figure*}
\centerline{\includegraphics[angle=-90,width=13cm]{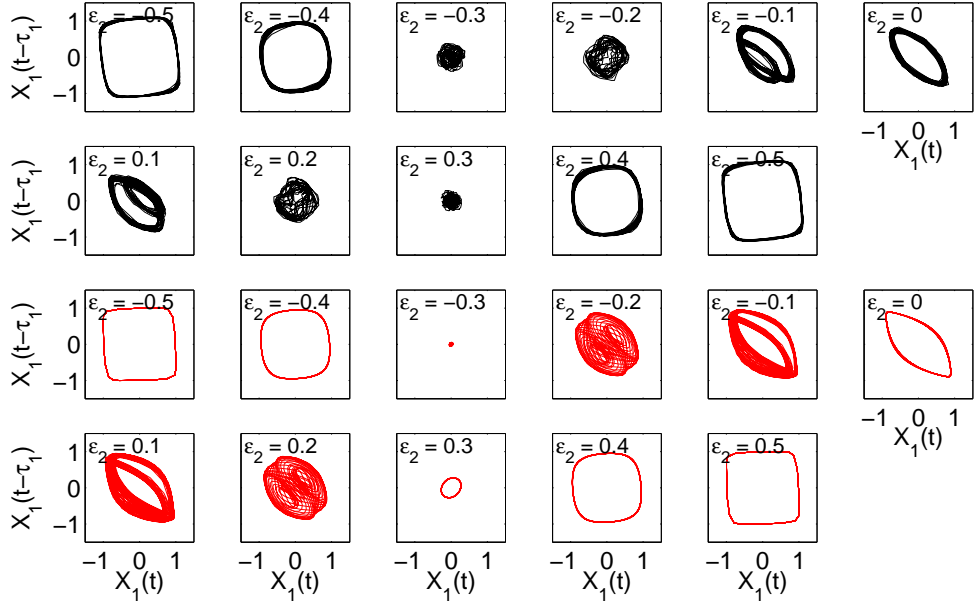}}
\caption{\label{oneosc} Time delay representations (phase portraits) of
the evolution of the mean field $X_1$ in dependence of the inter-group
coupling strength $\varepsilon_2$. Shown are the evolutions resulting
from the Langevin model (upper two rows, Eq. \ref{lang2c}) and the
mean field model (lower two rows, Eq. \ref{mf2c}). The parameters are:
$D=0.1$, $\tau_1=60$, $\tau_1=140$, $\varepsilon_1=-0.4$.}
\end{figure*}

In order to explore the dynamics of the system with two coupling
strengths we carry out numerical simulations of the Langevin model
(\ref{lang2c}) and compare the results with the theoretical
predictions derived in the previous Section. In these simulations the
strength of the intra-group coupling $\varepsilon_1$ and noise $D$ are
chosen so that in the absence of inter-group couplings
$\varepsilon_2=0$ the mean fields of the two oscillator groups $X_1$
and $X_2$ oscillate independently with frequencies $f_1\approx
1/2\tau_1$ and $f_2\approx 1/2\tau_2$ respectively
(cf. Sect.~\ref{dimod1}). We may then expect that for
$|\varepsilon_2|>0$ the system reveals dynamical properties
reminiscent of those of two coupled, limit-cycle oscillators, such as
chaotic behavior
\citep{Woafo96,Pastor93,Matthews90} and the amplitude death phenomenon
\citep{Atay03,Herrero00,Reddy98}. Fig.~\ref{oneosc} shows time
delay representations of the time series of $X_1(t)$ for inter-group
couplings of different strength $\varepsilon_2$. In certain regions
the system indeed shows the amplitude death phenomenon and in the
range $0<|\varepsilon|<\varepsilon_{\rm chaos}$ irregular motions are
observed. Numerical evidence suggests that these motions are chaotic.
Indeed, the time series analysis yields broad band power spectra and
positive maximum Lyapunov exponents. The determination of the Lyapunov
exponents is below discussed in greater detail.

\begin{figure}
\centerline{\includegraphics[angle=-90,width=8cm]{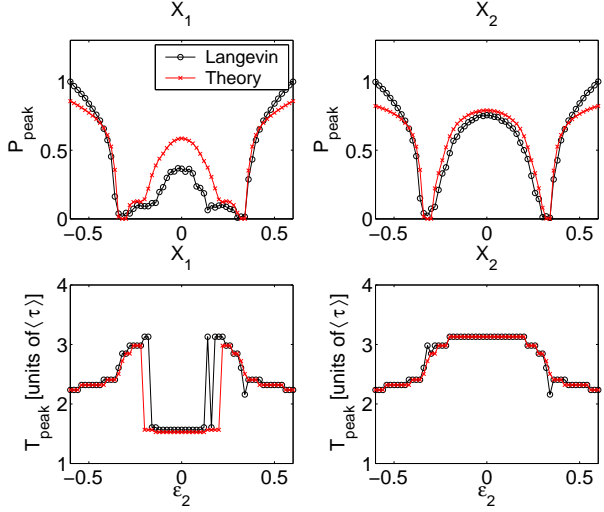}}
\caption{\label{veri2coup} The peak power $P_{\rm peak}$ (upper row) 
and the corresponding period $T_{\rm peak}=2\pi/\omega_{\rm peak}$
(lower row) of the two oscillator groups $X_1$ and $X_2$ resulting
from simulations of the theoretical mean field model (\ref{mf2c}) and
the Langevin model (\ref{lang2c}), respectively.  The parameters are
the same as in Fig.~\ref{oneosc}.}
\end{figure}

The comparison of the phase portraits in Fig.~\ref{oneosc} shows slight
deviations between theoretical predictions and the Langevin dynamics
(e.g. for $\varepsilon_2=-0.1$). These deviations stem from the
elimination of the noise fluctuations in the dichotomous model and
different phase shifts between the two oscillator groups $X_1$ and
$X_2$. However, the predictive power of our model is confirmed in
Fig.~\ref{veri2coup}, where the theoretical peak power $P_{\rm peak}$
and the corresponding period $T_{\rm peak}$ in dependence of the 
coupling strength $\varepsilon_1$ are compared with those resulting
from Langevin simulations.

Let us now explore the phase space of the system with nonuniform
couplings in greater detail.

The phase space regions of nontrivial stationary states are determined
by Eq.~(\ref{epsinterst}) and those where mean field oscillations and
amplitude death occur, are given by the roots of Eq.~(\ref{rootsof2ca})
and (\ref{rootsof2cb}). These roots are determined numerically. We
find that the solutions of Eq.~(\ref{omegaeps2a}) are a subset of the
the solutions of $f_{\rm r,i}(\omega)=0$. Thus, the corresponding 
critical values mark boundaries which qualitatively are equivalent
to those found in previous models. 

However, Eq.~(\ref{rootsof2ca}) also yields new solutions marking the
boundaries between the the zones of amplitude death and the areas of
nontrivial dynamics in the presence of weak inter-group
couplings. Within this areas there may occur islands of chaotic
dynamics. Indeed, an analysis of the mean field evolution yields
positive Lyapunov exponents for $0<|\varepsilon|<\varepsilon_{\rm
chaos}$. 

Since intrinsically our time delay system is infinite dimensional the
maximum Lyapunov exponents are here determined by an analysis of the
time series resulting from Eq.~(\ref{mf2c}). The analysis is carried
out using tools provided by the TISEAN software package
\citep{Hegger99,Kantz04}.


As stated above this process yields in some phase space regions clear
evidence of positive maximum Lyapunov exponents in the range
$0<\lambda\;[1/\bar{\tau}] \lesssim0.03$.

\begin{figure}
\centerline{\includegraphics[width=\hsize]{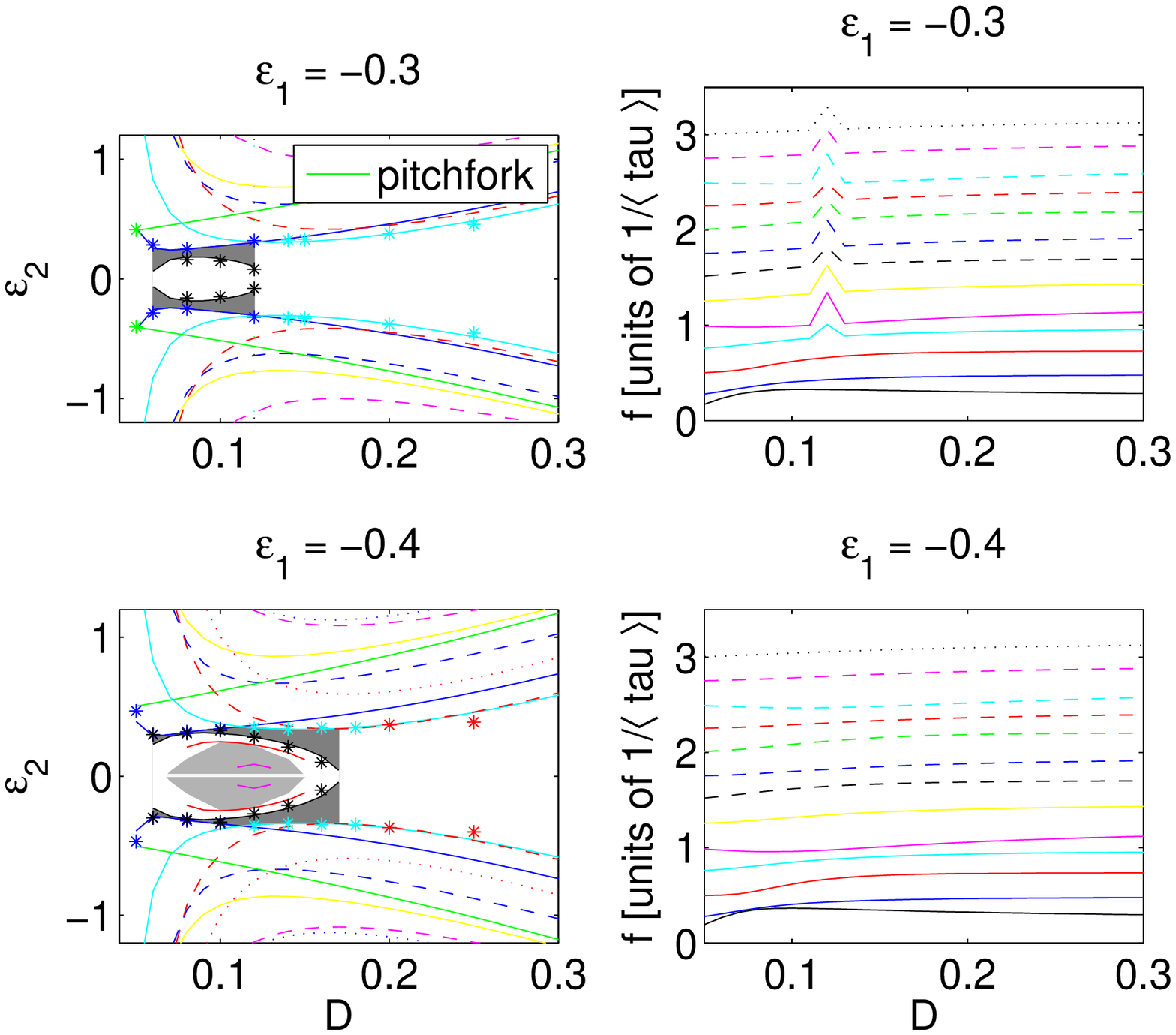}}
\caption{\label{phase2coup3} Same as in Fig.~\ref{multi2d} but this time
for a two time delay networks with nonuniform couplings (upper panels:
$\varepsilon=-0.3$, lower panels: $\varepsilon=-0.4$). Oscillatory
states as well as nontrivial stationary states occur for positive and
negative inter-group couplings $\varepsilon_2$. Dark gray areas depict
regions of amplitude death and the light gray areas mark regions of
chaotic dynamics. The time delays are $\tau_1=60$ and $\tau_2=140$.}
\end{figure}

The phase diagrams illustrating the different dynamic regions are
shown in Fig.~\ref{phase2coup3}. The Figure shows that chaotic
dynamics only occurs for strong intra-group
couplings $\varepsilon_1\gtrsim0.4$, i.e., when the individual
oscillations of the two groups are strong enough. 

\section{Summary and Conclusions}

The dynamics of networks of noisy bistable elements with time-delayed
couplings was studied analytically and numerically. Depending on the
noise level, the systems undergo ordering transitions and demonstrate
multistability; that is, for a strong enough positive feedback the
systems adopt a nonzero stationary mean field state, and a variety of
stable oscillatory mean field states are accessible for a positive as
well as negative feedback. The coherence of the oscillatory states is
maximal for a certain noise level; i.e., the systems demonstrate the
coherence resonance phenomenon.

For symmetric time delay distributions the frequencies of the
oscillations depend only on the mean time delay. However, the critical
couplings of the corresponding Hopf bifurcations depend also on the
higher order cumulants of the time delay distributions.  Indeed,
our findings suggest that nonuniformity of the time delays inhibits
the occurrence of the Hopf bifurcations and consequently increases the
stability of the trivial equilibrium. This may be important for time
delay systems such as neural networks and genetic regulatory networks,
since the degree of time delay nonuniformity, which is often related
to the diversity in the connectivity of the underlying network,
affects the accessibility of the nontrivial dynamical states.

The dichotomous theory based on delay-differential master-equations,
which has been developed in this article, adequately describes the
bifurcations of the trivial equilibrium in the limit of small noise
and coupling strength. Furthermore, the theory allows for the
application of a center manifold reduction and thus for a complete
bifurcation analysis of the trivial equilibrium. Far away from the
bifurcation points the mean field properties are well described by a
Gaussian approximation. However, a theoretical approach for the
description of the dynamics in the regime of strong noise near the
transition points is still lacking.

The collective dynamics of the networks of bistable elements with
uniform coupling strength is restricted to periodic oscillations and
stationary states. However, our model with nonuniform coupling
strengths shows that for certain coupling distributions, the system
behaves like a network of coupled limit cycle oscillators and
consequently, demonstrates in certain parameter-space areas the
amplitude death phenomenon or exhibits a chaotic evolution of the mean
field.

This paper discusses the dynamics of globally coupled systems with
time delays. However, in many systems the connectivity is sparse.
Since this is a particular case of systems with nonuniform coupling we
may expect that this endows the system with more complex dynamical
properties. This issue should be addressed in future studies.

\section*{Acknowledgements}

We thank A. Pikovsky for many valuable discussions. This work was
supported by the Swiss National Science Foundation (D.H.) and by the
U.S. Department of Energy, Office of Basic Energy Sciences under grant
DE-FG-03-96ER14592 (L.T.).


\begin{thebibliography}{46}
\expandafter\ifx\csname natexlab\endcsname\relax\def\natexlab#1{#1}\fi
\expandafter\ifx\csname bibnamefont\endcsname\relax
  \def\bibnamefont#1{#1}\fi
\expandafter\ifx\csname bibfnamefont\endcsname\relax
  \def\bibfnamefont#1{#1}\fi
\expandafter\ifx\csname citenamefont\endcsname\relax
  \def\citenamefont#1{#1}\fi
\expandafter\ifx\csname url\endcsname\relax
  \def\url#1{\texttt{#1}}\fi
\expandafter\ifx\csname urlprefix\endcsname\relax\def\urlprefix{URL }\fi
\providecommand{\bibinfo}[2]{#2}
\providecommand{\eprint}[2][]{\url{#2}}

\bibitem[{\citenamefont{{Dawson}}(1983)}]{Dawson83}
\bibinfo{author}{\bibfnamefont{D.}~\bibnamefont{{Dawson}}},
  \bibinfo{journal}{J. Stat. Phys.} \textbf{\bibinfo{volume}{29}},
  \bibinfo{pages}{31} (\bibinfo{year}{1983}).

\bibitem[{\citenamefont{{Jung} et~al.}(1992)\citenamefont{{Jung}, {Behn},
  {Pantazelou}, and {Moss}}}]{Jung92}
\bibinfo{author}{\bibfnamefont{P.}~\bibnamefont{{Jung}}},
  \bibinfo{author}{\bibfnamefont{U.}~\bibnamefont{{Behn}}},
  \bibinfo{author}{\bibfnamefont{E.}~\bibnamefont{{Pantazelou}}},
  \bibnamefont{and} \bibinfo{author}{\bibfnamefont{F.}~\bibnamefont{{Moss}}},
  \bibinfo{journal}{Phys. Rev. A} \textbf{\bibinfo{volume}{46}},
  \bibinfo{pages}{R1709} (\bibinfo{year}{1992}).

\bibitem[{\citenamefont{{Koulakov} et~al.}(2002)\citenamefont{{Koulakov},
  {Raghavachari}, {Kepecs}, and {Lisman}}}]{Koulakov02}
\bibinfo{author}{\bibfnamefont{A.}~\bibnamefont{{Koulakov}}},
  \bibinfo{author}{\bibfnamefont{S.}~\bibnamefont{{Raghavachari}}},
  \bibinfo{author}{\bibfnamefont{A.}~\bibnamefont{{Kepecs}}}, \bibnamefont{and}
  \bibinfo{author}{\bibfnamefont{J.}~\bibnamefont{{Lisman}}},
  \bibinfo{journal}{Nat. Neurosci.} \textbf{\bibinfo{volume}{5}},
  \bibinfo{pages}{775} (\bibinfo{year}{2002}).

\bibitem[{\citenamefont{{Camperi} and {Wang}}(1998)}]{Camperi98}
\bibinfo{author}{\bibfnamefont{M.}~\bibnamefont{{Camperi}}} \bibnamefont{and}
  \bibinfo{author}{\bibfnamefont{X.}~\bibnamefont{{Wang}}},
  \bibinfo{journal}{J. Comp. Neurosci.} \textbf{\bibinfo{volume}{5}},
  \bibinfo{pages}{383} (\bibinfo{year}{1998}).

\bibitem[{\citenamefont{{Sompolinsky}}(1986)}]{Sompolinsky86}
\bibinfo{author}{\bibfnamefont{H.}~\bibnamefont{{Sompolinsky}}},
  \bibinfo{journal}{Phys. Rev. A} \textbf{\bibinfo{volume}{34}},
  \bibinfo{pages}{2571} (\bibinfo{year}{1986}).

\bibitem[{\citenamefont{{Gardner} et~al.}(2000)\citenamefont{{Gardner},
  {Cantor}, and {Collins}}}]{Gardner00}
\bibinfo{author}{\bibfnamefont{T.}~\bibnamefont{{Gardner}}},
  \bibinfo{author}{\bibfnamefont{C.}~\bibnamefont{{Cantor}}}, \bibnamefont{and}
  \bibinfo{author}{\bibfnamefont{J.}~\bibnamefont{{Collins}}},
  \bibinfo{journal}{Nature} \textbf{\bibinfo{volume}{403}},
  \bibinfo{pages}{339} (\bibinfo{year}{2000}).

\bibitem[{\citenamefont{{Zanette}}(1997)}]{Zanette97}
\bibinfo{author}{\bibfnamefont{D.~H.} \bibnamefont{{Zanette}}},
  \bibinfo{journal}{Phys. Rev. E} \textbf{\bibinfo{volume}{55}},
  \bibinfo{pages}{5315} (\bibinfo{year}{1997}).

\bibitem[{\citenamefont{{Reddy} et~al.}(1998)\citenamefont{{Reddy}, {Sen}, and
  {Johnston}}}]{Reddy98}
\bibinfo{author}{\bibfnamefont{D.}~\bibnamefont{{Reddy}}},
  \bibinfo{author}{\bibfnamefont{A.}~\bibnamefont{{Sen}}}, \bibnamefont{and}
  \bibinfo{author}{\bibfnamefont{G.}~\bibnamefont{{Johnston}}},
  \bibinfo{journal}{Phys. Rev. Lett.} \textbf{\bibinfo{volume}{80}},
  \bibinfo{pages}{5109} (\bibinfo{year}{1998}).

\bibitem[{\citenamefont{{Nakamura} et~al.}(1994)\citenamefont{{Nakamura},
  {Tominaga}, and {Munakata}}}]{Nakamura94}
\bibinfo{author}{\bibfnamefont{Y.}~\bibnamefont{{Nakamura}}},
  \bibinfo{author}{\bibfnamefont{F.}~\bibnamefont{{Tominaga}}},
  \bibnamefont{and}
  \bibinfo{author}{\bibfnamefont{T.}~\bibnamefont{{Munakata}}},
  \bibinfo{journal}{Phys. Rev. E} \textbf{\bibinfo{volume}{49}},
  \bibinfo{pages}{4849} (\bibinfo{year}{1994}).

\bibitem[{\citenamefont{{Bressloff} and {Coombes}}(1998)}]{Bresseloff98}
\bibinfo{author}{\bibfnamefont{P.~C.} \bibnamefont{{Bressloff}}}
  \bibnamefont{and}
  \bibinfo{author}{\bibfnamefont{S.}~\bibnamefont{{Coombes}}},
  \bibinfo{journal}{Phys. Rev. Lett.} \textbf{\bibinfo{volume}{80}},
  \bibinfo{pages}{4815} (\bibinfo{year}{1998}).

\bibitem[{\citenamefont{{Choi} and {Huberman}}(1985)}]{Choi85}
\bibinfo{author}{\bibfnamefont{M.~Y.} \bibnamefont{{Choi}}} \bibnamefont{and}
  \bibinfo{author}{\bibfnamefont{B.~A.} \bibnamefont{{Huberman}}},
  \bibinfo{journal}{Phys. Rev. B} \textbf{\bibinfo{volume}{31}},
  \bibinfo{pages}{2862} (\bibinfo{year}{1985}).

\bibitem[{\citenamefont{{Wiesenfeld} et~al.}(1990)\citenamefont{{Wiesenfeld},
  {Bracikowski}, {James}, and {Roy}}}]{Wiesenfeld90}
\bibinfo{author}{\bibfnamefont{K.}~\bibnamefont{{Wiesenfeld}}},
  \bibinfo{author}{\bibfnamefont{C.}~\bibnamefont{{Bracikowski}}},
  \bibinfo{author}{\bibfnamefont{G.}~\bibnamefont{{James}}}, \bibnamefont{and}
  \bibinfo{author}{\bibfnamefont{R.}~\bibnamefont{{Roy}}},
  \bibinfo{journal}{Phys. Rev. Lett.} \textbf{\bibinfo{volume}{65}},
  \bibinfo{pages}{1749} (\bibinfo{year}{1990}).

\bibitem[{\citenamefont{{Wiesenfeld} and {Hadley}}(1989)}]{Wiesenfeld89}
\bibinfo{author}{\bibfnamefont{K.}~\bibnamefont{{Wiesenfeld}}}
  \bibnamefont{and} \bibinfo{author}{\bibfnamefont{P.}~\bibnamefont{{Hadley}}},
  \bibinfo{journal}{Phys. Rev. Lett.} \textbf{\bibinfo{volume}{62}},
  \bibinfo{pages}{1335} (\bibinfo{year}{1989}).

\bibitem[{\citenamefont{{Yeung} and {Strogatz}}(1999)}]{Yeung99}
\bibinfo{author}{\bibfnamefont{M.~K.~S.} \bibnamefont{{Yeung}}}
  \bibnamefont{and} \bibinfo{author}{\bibfnamefont{S.~H.}
  \bibnamefont{{Strogatz}}}, \bibinfo{journal}{Phys. Rev. Lett.}
  \textbf{\bibinfo{volume}{82}}, \bibinfo{pages}{648} (\bibinfo{year}{1999}).

\bibitem[{\citenamefont{{Van den Broeck} and {Parrondo}}(1994)}]{Broeck94}
\bibinfo{author}{\bibfnamefont{C.}~\bibnamefont{{Van den Broeck}}}
  \bibnamefont{and}
  \bibinfo{author}{\bibfnamefont{J.}~\bibnamefont{{Parrondo}}},
  \bibinfo{journal}{Phys. Rev. E} \textbf{\bibinfo{volume}{49}},
  \bibinfo{pages}{2639} (\bibinfo{year}{1994}).

\bibitem[{\citenamefont{{Kuramoto}}(1991)}]{Kuramoto91}
\bibinfo{author}{\bibfnamefont{Y.}~\bibnamefont{{Kuramoto}}},
  \emph{\bibinfo{title}{Chemical Oscillations, Waves and Turbulence}}
  (\bibinfo{publisher}{Springer}, \bibinfo{address}{Berlin},
  \bibinfo{year}{1991}).

\bibitem[{\citenamefont{{Shiino}}(1987)}]{Shiino87}
\bibinfo{author}{\bibfnamefont{M.}~\bibnamefont{{Shiino}}},
  \bibinfo{journal}{Phys. Rev. A} \textbf{\bibinfo{volume}{36}},
  \bibinfo{pages}{2393} (\bibinfo{year}{1987}).

\bibitem[{\citenamefont{{Desai} and {Zwanzig}}(1978)}]{Desai78}
\bibinfo{author}{\bibfnamefont{R.~C.} \bibnamefont{{Desai}}} \bibnamefont{and}
  \bibinfo{author}{\bibfnamefont{R.}~\bibnamefont{{Zwanzig}}},
  \bibinfo{journal}{J. Stat. Phys.} \textbf{\bibinfo{volume}{19}},
  \bibinfo{pages}{1} (\bibinfo{year}{1978}).

\bibitem[{\citenamefont{{Tsimring} and {Pikovsky}}(2001)}]{Tsimring01}
\bibinfo{author}{\bibfnamefont{L.~S.} \bibnamefont{{Tsimring}}}
  \bibnamefont{and}
  \bibinfo{author}{\bibfnamefont{A.}~\bibnamefont{{Pikovsky}}},
  \bibinfo{journal}{Phys. Rev. Lett.} \textbf{\bibinfo{volume}{87}},
  \bibinfo{pages}{250602} (\bibinfo{year}{2001}).

\bibitem[{\citenamefont{{Huber} and {Tsimring}}(2003)}]{Huber03}
\bibinfo{author}{\bibfnamefont{D.}~\bibnamefont{{Huber}}} \bibnamefont{and}
  \bibinfo{author}{\bibfnamefont{L.~S.} \bibnamefont{{Tsimring}}},
  \bibinfo{journal}{Phys. Rev. Lett.} \textbf{\bibinfo{volume}{91}},
  \bibinfo{pages}{260601} (\bibinfo{year}{2003}).

\bibitem[{\citenamefont{{Salami} et~al.}(2003)\citenamefont{{Salami}, {Itami},
  {Tsumoto}, and {Kimura}}}]{Salami03}
\bibinfo{author}{\bibfnamefont{M.}~\bibnamefont{{Salami}}},
  \bibinfo{author}{\bibfnamefont{C.}~\bibnamefont{{Itami}}},
  \bibinfo{author}{\bibfnamefont{T.}~\bibnamefont{{Tsumoto}}},
  \bibnamefont{and} \bibinfo{author}{\bibfnamefont{F.}~\bibnamefont{{Kimura}}},
  \bibinfo{journal}{PNAS} \textbf{\bibinfo{volume}{100}}, \bibinfo{pages}{6174}
  (\bibinfo{year}{2003}).

\bibitem[{\citenamefont{{Paulsson} and {Ehrenberg}}(2001)}]{Paulsson01}
\bibinfo{author}{\bibfnamefont{J.}~\bibnamefont{{Paulsson}}} \bibnamefont{and}
  \bibinfo{author}{\bibfnamefont{M.}~\bibnamefont{{Ehrenberg}}},
  \bibinfo{journal}{Q. Rev. Biophys.} \textbf{\bibinfo{volume}{34}},
  \bibinfo{pages}{1} (\bibinfo{year}{2001}).

\bibitem[{\citenamefont{{Atay}}(2003)}]{Atay03}
\bibinfo{author}{\bibfnamefont{F.}~\bibnamefont{{Atay}}},
  \bibinfo{journal}{Phys. Rev. Lett.} \textbf{\bibinfo{volume}{91}},
  \bibinfo{pages}{094101} (\bibinfo{year}{2003}).

\bibitem[{\citenamefont{{H{\" a}nggi} et~al.}(1990)\citenamefont{{H{\" a}nggi},
  {Talkner}, and {Borkovec}}}]{Haenggi90}
\bibinfo{author}{\bibfnamefont{P.}~\bibnamefont{{H{\" a}nggi}}},
  \bibinfo{author}{\bibfnamefont{P.}~\bibnamefont{{Talkner}}},
  \bibnamefont{and}
  \bibinfo{author}{\bibfnamefont{M.}~\bibnamefont{{Borkovec}}},
  \bibinfo{journal}{Rev. Mod. Phys.} \textbf{\bibinfo{volume}{62}},
  \bibinfo{pages}{251} (\bibinfo{year}{1990}).

\bibitem[{\citenamefont{{Kramers}}(1940)}]{Kramers40}
\bibinfo{author}{\bibfnamefont{H.}~\bibnamefont{{Kramers}}},
  \bibinfo{journal}{Physica (Utrecht)} \textbf{\bibinfo{volume}{7}},
  \bibinfo{pages}{284} (\bibinfo{year}{1940}).

\bibitem[{\citenamefont{{Pikovsky} et~al.}(2002)\citenamefont{{Pikovsky},
  {Zaikin}, and {de La Casa}}}]{Pikovsky02}
\bibinfo{author}{\bibfnamefont{A.}~\bibnamefont{{Pikovsky}}},
  \bibinfo{author}{\bibfnamefont{A.}~\bibnamefont{{Zaikin}}}, \bibnamefont{and}
  \bibinfo{author}{\bibfnamefont{M.~A.} \bibnamefont{{de La Casa}}},
  \bibinfo{journal}{Phys. Rev. Lett.} \textbf{\bibinfo{volume}{88}},
  \bibinfo{pages}{050601} (\bibinfo{year}{2002}).

\bibitem[{\citenamefont{{McNamara} and {Wiesenfeld}}(1989)}]{McNamara89}
\bibinfo{author}{\bibfnamefont{B.}~\bibnamefont{{McNamara}}} \bibnamefont{and}
  \bibinfo{author}{\bibfnamefont{K.}~\bibnamefont{{Wiesenfeld}}},
  \bibinfo{journal}{Phys. Rev. A} \textbf{\bibinfo{volume}{39}},
  \bibinfo{pages}{4854} (\bibinfo{year}{1989}).

\bibitem[{\citenamefont{{Gammaitoni} et~al.}(1998)\citenamefont{{Gammaitoni},
  {H{\" a}nggi}, {Jung}, and {Marchesoni}}}]{Gammaitoni98}
\bibinfo{author}{\bibfnamefont{L.}~\bibnamefont{{Gammaitoni}}},
  \bibinfo{author}{\bibfnamefont{P.}~\bibnamefont{{H{\" a}nggi}}},
  \bibinfo{author}{\bibfnamefont{P.}~\bibnamefont{{Jung}}}, \bibnamefont{and}
  \bibinfo{author}{\bibfnamefont{F.}~\bibnamefont{{Marchesoni}}},
  \bibinfo{journal}{Rev. Mod. Phys.} \textbf{\bibinfo{volume}{70}},
  \bibinfo{pages}{223} (\bibinfo{year}{1998}).

\bibitem[{\citenamefont{{Chow} and {Hale}}(1982)}]{Chow82}
\bibinfo{author}{\bibfnamefont{S.~N.} \bibnamefont{{Chow}}} \bibnamefont{and}
  \bibinfo{author}{\bibfnamefont{J.~K.} \bibnamefont{{Hale}}},
  \emph{\bibinfo{title}{Methods of Bifurcation Theory}}
  (\bibinfo{publisher}{Springer}, \bibinfo{address}{New York},
  \bibinfo{year}{1982}).

\bibitem[{\citenamefont{{Faria} and {Magalh\~aes}}(1995)}]{Faria95}
\bibinfo{author}{\bibfnamefont{T.}~\bibnamefont{{Faria}}} \bibnamefont{and}
  \bibinfo{author}{\bibfnamefont{L.~T.} \bibnamefont{{Magalh\~aes}}},
  \bibinfo{journal}{J . Diff. Eq.} \textbf{\bibinfo{volume}{122}},
  \bibinfo{pages}{181} (\bibinfo{year}{1995}).

\bibitem[{\citenamefont{{Giannakopoulos} and {Zapp}}(1999)}]{Giannakopoulos99}
\bibinfo{author}{\bibfnamefont{F.}~\bibnamefont{{Giannakopoulos}}}
  \bibnamefont{and} \bibinfo{author}{\bibfnamefont{A.}~\bibnamefont{{Zapp}}},
  \bibinfo{journal}{J. Math. Anal. Appl.} \textbf{\bibinfo{volume}{237}},
  \bibinfo{pages}{425} (\bibinfo{year}{1999}).

\bibitem[{\citenamefont{{Redmond} et~al.}(2002)\citenamefont{{Redmond},
  {LeBlanc}, and {Longtin}}}]{Redmond02}
\bibinfo{author}{\bibfnamefont{B.~F.} \bibnamefont{{Redmond}}},
  \bibinfo{author}{\bibfnamefont{V.~G.} \bibnamefont{{LeBlanc}}},
  \bibnamefont{and}
  \bibinfo{author}{\bibfnamefont{A.}~\bibnamefont{{Longtin}}},
  \bibinfo{journal}{Physica D} \textbf{\bibinfo{volume}{166}},
  \bibinfo{pages}{131} (\bibinfo{year}{2002}).

\bibitem[{\citenamefont{{Strogatz}}(2000)}]{Strogatz}
\bibinfo{author}{\bibfnamefont{S.~H.} \bibnamefont{{Strogatz}}},
  \emph{\bibinfo{title}{Nonlinear Dynamics and Chaos}}
  (\bibinfo{publisher}{Westview Press}, \bibinfo{address}{Cambridge},
  \bibinfo{year}{2000}).

\bibitem[{\citenamefont{{Kiss} et~al.}(2003)\citenamefont{{Kiss}, {Hudson},
  {Escalera Santos}, and {Parmananda}}}]{Kiss03}
\bibinfo{author}{\bibfnamefont{I.~Z.} \bibnamefont{{Kiss}}},
  \bibinfo{author}{\bibfnamefont{J.~L.} \bibnamefont{{Hudson}}},
  \bibinfo{author}{\bibfnamefont{G.~J.} \bibnamefont{{Escalera Santos}}},
  \bibnamefont{and}
  \bibinfo{author}{\bibfnamefont{P.}~\bibnamefont{{Parmananda}}},
  \bibinfo{journal}{Phys. Rev. E} \textbf{\bibinfo{volume}{67}},
  \bibinfo{pages}{035201} (\bibinfo{year}{2003}).

\bibitem[{\citenamefont{{Miyakawa} and {Isikawa}}(2002)}]{Miyakawa02}
\bibinfo{author}{\bibfnamefont{K.}~\bibnamefont{{Miyakawa}}} \bibnamefont{and}
  \bibinfo{author}{\bibfnamefont{H.}~\bibnamefont{{Isikawa}}},
  \bibinfo{journal}{Phys. Rev. E} \textbf{\bibinfo{volume}{66}},
  \bibinfo{pages}{046204} (\bibinfo{year}{2002}).

\bibitem[{\citenamefont{{Pikovsky} and {Kurths}}(1997)}]{Pikovsky97}
\bibinfo{author}{\bibfnamefont{A.~S.} \bibnamefont{{Pikovsky}}}
  \bibnamefont{and} \bibinfo{author}{\bibfnamefont{J.}~\bibnamefont{{Kurths}}},
  \bibinfo{journal}{Phys. Rev. Lett.} \textbf{\bibinfo{volume}{78}},
  \bibinfo{pages}{775} (\bibinfo{year}{1997}).

\bibitem[{\citenamefont{{Sigeti} and {Horsthemke}}(1989)}]{Sigeti89}
\bibinfo{author}{\bibfnamefont{D.}~\bibnamefont{{Sigeti}}} \bibnamefont{and}
  \bibinfo{author}{\bibfnamefont{W.}~\bibnamefont{{Horsthemke}}},
  \bibinfo{journal}{J. Stat. Phys.} \textbf{\bibinfo{volume}{54}},
  \bibinfo{pages}{1217} (\bibinfo{year}{1989}).

\bibitem[{\citenamefont{{Zhou} et~al.}(2001)\citenamefont{{Zhou}, {Kurths}, and
  {Hu}}}]{Zhou01}
\bibinfo{author}{\bibfnamefont{C.}~\bibnamefont{{Zhou}}},
  \bibinfo{author}{\bibfnamefont{J.}~\bibnamefont{{Kurths}}}, \bibnamefont{and}
  \bibinfo{author}{\bibfnamefont{B.}~\bibnamefont{{Hu}}},
  \bibinfo{journal}{Phys. Rev. Lett.} \textbf{\bibinfo{volume}{87}},
  \bibinfo{pages}{98101} (\bibinfo{year}{2001}).

\bibitem[{\citenamefont{{Van Kampen}}(2003)}]{Kampen03}
\bibinfo{author}{\bibfnamefont{N.}~\bibnamefont{{Van Kampen}}},
  \emph{\bibinfo{title}{Stochastic Processes in Physics and Chemistry}}
  (\bibinfo{publisher}{Elesiver Scinece B.V.}, \bibinfo{address}{Amsterdam, The
  Netherlands}, \bibinfo{year}{2003}).

\bibitem[{\citenamefont{{Risken}}(1989)}]{Risken89}
\bibinfo{author}{\bibfnamefont{H.}~\bibnamefont{{Risken}}},
  \emph{\bibinfo{title}{The Fokker-Planck Equation}}
  (\bibinfo{publisher}{Springer}, \bibinfo{address}{Berlin Heidelberg},
  \bibinfo{year}{1989}).

\bibitem[{\citenamefont{{Woafo} et~al.}(1996)\citenamefont{{Woafo}, {Chedjou},
  and {Fotsin}}}]{Woafo96}
\bibinfo{author}{\bibfnamefont{P.}~\bibnamefont{{Woafo}}},
  \bibinfo{author}{\bibfnamefont{J.~C.} \bibnamefont{{Chedjou}}},
  \bibnamefont{and} \bibinfo{author}{\bibfnamefont{H.~B.}
  \bibnamefont{{Fotsin}}}, \bibinfo{journal}{Phys. Rev. E}
  \textbf{\bibinfo{volume}{54}}, \bibinfo{pages}{5929} (\bibinfo{year}{1996}).

\bibitem[{\citenamefont{{Pastor} et~al.}(1993)\citenamefont{{Pastor}, {P{\'
  e}rez-Garc{\'{\i}}a}, {Encinas}, and {Guerra}}}]{Pastor93}
\bibinfo{author}{\bibfnamefont{I.}~\bibnamefont{{Pastor}}},
  \bibinfo{author}{\bibfnamefont{V.~M.} \bibnamefont{{P{\'
  e}rez-Garc{\'{\i}}a}}},
  \bibinfo{author}{\bibfnamefont{F.}~\bibnamefont{{Encinas}}},
  \bibnamefont{and} \bibinfo{author}{\bibfnamefont{J.~M.}
  \bibnamefont{{Guerra}}}, \bibinfo{journal}{Phys. Rev. E}
  \textbf{\bibinfo{volume}{48}}, \bibinfo{pages}{171} (\bibinfo{year}{1993}).

\bibitem[{\citenamefont{{Matthews} and {Strogatz}}(1990)}]{Matthews90}
\bibinfo{author}{\bibfnamefont{P.~C.} \bibnamefont{{Matthews}}}
  \bibnamefont{and} \bibinfo{author}{\bibfnamefont{S.~H.}
  \bibnamefont{{Strogatz}}}, \bibinfo{journal}{Phys. Rev. Lett.}
  \textbf{\bibinfo{volume}{65}}, \bibinfo{pages}{1701} (\bibinfo{year}{1990}).

\bibitem[{\citenamefont{{Herrero} et~al.}(2000)\citenamefont{{Herrero},
  {Figueras}, {Rius}, {Pi}, and {Orriols}}}]{Herrero00}
\bibinfo{author}{\bibfnamefont{R.}~\bibnamefont{{Herrero}}},
  \bibinfo{author}{\bibfnamefont{M.}~\bibnamefont{{Figueras}}},
  \bibinfo{author}{\bibfnamefont{J.}~\bibnamefont{{Rius}}},
  \bibinfo{author}{\bibfnamefont{F.}~\bibnamefont{{Pi}}}, \bibnamefont{and}
  \bibinfo{author}{\bibfnamefont{G.}~\bibnamefont{{Orriols}}},
  \bibinfo{journal}{Phys. Rev. Lett.} \textbf{\bibinfo{volume}{84}},
  \bibinfo{pages}{5312} (\bibinfo{year}{2000}).

\bibitem[{\citenamefont{{Hegger} et~al.}(1999)\citenamefont{{Hegger}, {Kantz},
  and {Schreiber}}}]{Hegger99}
\bibinfo{author}{\bibfnamefont{R.}~\bibnamefont{{Hegger}}},
  \bibinfo{author}{\bibfnamefont{H.}~\bibnamefont{{Kantz}}}, \bibnamefont{and}
  \bibinfo{author}{\bibfnamefont{T.}~\bibnamefont{{Schreiber}}},
  \bibinfo{journal}{Chaos} \textbf{\bibinfo{volume}{9}}, \bibinfo{pages}{413}
  (\bibinfo{year}{1999}).

\bibitem[{\citenamefont{{Kantz}}(2004)}]{Kantz04}
\bibinfo{author}{\bibfnamefont{H.}~\bibnamefont{{Kantz}}},
  \emph{\bibinfo{title}{Nonlinear Time Series Analysis}}
  (\bibinfo{publisher}{Cambridge University Press},
  \bibinfo{address}{Cambridge}, \bibinfo{year}{2004}).

\end{thebibliography}

\end{document}